\pdfoutput=1

\documentclass[aps,prb,reprint,superscriptaddress,longbibliography]{revtex4-2}

\usepackage[utf8]{inputenc}
\usepackage{bm}
\usepackage{siunitx}

\usepackage{enumerate}
\usepackage{amsfonts}
\usepackage{amsmath}
\usepackage{amssymb}
\usepackage{color}
\usepackage{soul}
\usepackage{verbatim}

\usepackage{graphicx}

\usepackage{xurl}
\usepackage[colorlinks,allcolors=blue]{hyperref}

\usepackage[capitalize]{cleveref}

\definecolor{DarkRed}{rgb}{0.80,0,0}
\definecolor{DarkGray}{rgb}{0.7,0.7,0.7}

\newcommand{\prlsection}[1]{\textit{#1}.\kern0.05em---\kern0.05em\ignorespaces}

\begin{document}
\title{Magnonic spin Joule heating and rectification effects}
\author{Morten Amundsen}
\affiliation{Nordita,
KTH Royal Institute of Technology and Stockholm University,
Hannes Alfvéns väg 12, SE-106 91 Stockholm, Sweden}
\author{Irina V. Bobkova}
\affiliation{Institute of Solid State Physics, 
Chernogolovka, 142432 Moscow , Russia}
\affiliation{Moscow Institute of Physics and Technology, 
Dolgoprudny, 141700 Moscow, Russia}
\affiliation{National Research University Higher School of Economics, Moscow, 101000 Russia}
\author{Akashdeep Kamra}
\affiliation{Condensed Matter Physics Center (IFIMAC) and Departamento de Física Teórica de la Materia Condensada, Universidad Autónoma de Madrid, E-28049 Madrid, Spain }
\begin{abstract}
	Nonlinear devices, such as transistors, enable contemporary computing technologies. We theoretically investigate nonlinear effects, bearing a high fundamental scientific and technical relevance, in magnonics with emphasis on superconductor-ferromagnet hybrids. Accounting for finite magnon chemical potential, we theoretically demonstrate magnonic spin-Joule heating, the spin analogue of conventional electronic Joule heating. Besides suggesting a key contribution to magnonic heat transport in a broad range of devices, it provides insights into the thermal physics of non-conserved bosonic excitations. Considering a spin-split superconductor self-consistently, we demonstrate its interface with a ferromagnetic insulator to harbor large tunability of spin and thermal conductances. We further demonstrate hysteretic rectification I-V characteristics in this hybrid, where the hysteresis results from the superconducting state bistability.
\end{abstract}
\maketitle

\section{Introduction}
The spin carriers in magnetic insulators - magnons - constitute a fertile platform for science and technology due to their nonconserved bosonic nature as well as solid-state host that admits strong interactions. For example, creating nonequilibrium magnons, allowed due to their general lack of conservation, together with strong magnon-magnon scattering enables formation of bosonic condensates~\cite{Demokritov2006,Demidov2008,Bender2012}, not feasible with fermionic electrons. Since they transport information without a movement of electrons, and Joule heating due to charge current flow, they are also touted as low-dissipation alternatives to electrons as information carriers~\cite{Bauer2012,Kruglyak2010,Serga2010,Chumak2015,Chumak2017,Althammer2018,Althammer2021,Pirro2021,Barman2021,Nakata2017}. While magnon chemical potential vanishes in equilibrium due to their nonconserved nature, it becomes nonzero in nonequilibrium situations due to their almost conserved nature at short time scales. This underlies a vast range of phenomena such as Bose-Einstein condensation~\cite{Demokritov2006,Demidov2008,Bender2012} and spin transport~\cite{Cornelissen2016,Kamra2016,Olsson2020,Schlitz2021}, where the importance of nonzero chemical potential in these phenomena has recently been, and continues to be, recognized~\cite{Du2017,Cornelissen2016}. Overall, a broad range of magnonic devices have already been demonstrated~\cite{Kajiwara2010a,Cornelissen2015,Goennenwein2015,Shan2017,Cornelissen2018,Wimmer2019,Wimmer2020,An2021,Schlitz2021}.

From the perspective of both devices and exciting physics, nonlinearities are highly desired, whether for unconventional computing paradigms~\cite{Markovic2020} or Maxwell's daemon-like switches~\cite{Kish2012,Koski2016}. In this context, the high potential of synergy between magnonics and superconductors has just begun to be realized~\cite{Bergeret2018,Heikkila2019,Yang2021,Jeon2020}. The latter admit strong nonlinear effects since a new and small energy scale - the superconducting gap - determines properties such as quasiparticle density of states~\cite{Strambini2017,Machon2013,Machon2014,Ozaeta2014}. Recent years have seen an upsurge of activity in this context with several exciting phenomena discussed, experimentally~\cite{Wolf2014,Golovchanskiy2018,Dobrovolskiy2019,Li2019,Jeon2019,Lachance-Quirion2020,Jeon2020,Jeon2020a,Golovchanskiy2020,Golovchanskiy2021} as well as theoretically~\cite{Machon2013,Machon2014,Ozaeta2014,Ohnuma2017,Kato2019,Chakraborty2019,Vargas2020,Vargas2020a,Ojajarvi2021,Ahari2021}. The possibility of transport influencing, and even destroying, the superconducting state may be considered as the pinnacle of nonlinearity offered by a superconductor and has been exploited successfully in various charge transport based devices~\cite{Bardeen1962,Moraru2006,Li2013}. 

Here, we theoretically investigate spin and heat transport in hybrids consisting of a superconductor (SC) interfaced with a ferromagnetic insulator (FI). Our focus is on exploiting the strong nonlinearities available in this hybrid for new device concepts and fundamental physics. Hence, we evaluate the superconducting state self-consistently. Our first key finding is that magnonic spin current flow results in a spin-Joule heating given by $I_m^2 R_m$, where $I_m$ is the magnon current and $R_m$ is the spin resistance. This finding is not specific to the hybrids considered and bears relevance for magnon chemical potential driven spin transport in general. In bulk magnets, the spin-Joule heating power per unit volume becomes $\pmb{j}_m^2 \rho_m$. Here, $\pmb{j}_m$ is the magnon current density and $\rho_m$, the spin resistivity, is defined via $\pmb{j}_m = - \pmb{\nabla} \mu_m / \rho_m$, where $\mu_m$ is the magnon chemical potential~\cite{Cornelissen2016}. Furthermore, accounting for dipolar interactions that do not conserve spin, we find new contributions to spin Joule heating that are absent in conventional Joule heating. Our second set of findings demonstrates a control over the transport coefficients, such as spin and thermal conductances, in the SC-FI interface via spin-splitting in the SC layer. Further, via self-consistent calculations, we demonstrate a hysteretic rectification effect in the spin current vs. spin chemical potential difference. This is rooted in the dependence of the superconducting state on the spin chemical potential, as the latter results in depairing of spin-singlet Cooper pairs~\cite{Bergeret2018}. Besides the interesting directionality, such an ``I-V'' characteristic could enable devices with inbuilt memory and threshold behavior, beneficial in some unconventional computing architectures~\cite{Markovic2020}.

\begin{figure}[tb]
\includegraphics[width=\columnwidth]{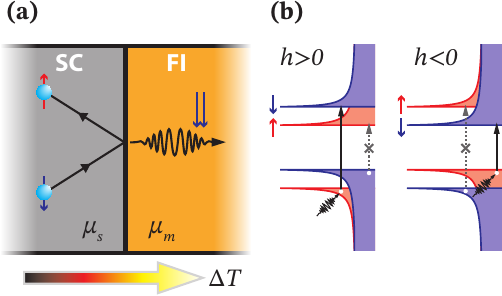}
\caption{In (a) is shown a sketch of the system under consideration. A Zeeman split superconductor is placed in contact with a ferromagnetic insulator. (b) illustrates the physical picture, for a case where a magnon is absorbed at the interface. Since the magnon has a spin of $-\hbar$, only a spin-flip process in which a spin up ($+\hbar/2$) quasiparticle is converted to spin down ($-\hbar/2$) is then allowed. Due to the Zeeman splitting, the gap for such a process to occur is higher when the exchange field $h$ is parallel to the magnetization of the ferromagnetic insulator, and vice versa.}
\label{fig:model}
\end{figure}

\section{Spin-Joule heating}
Considering an interface between an FI and SC [Fig.~\ref{fig:model} (a)], the heat generated per unit time in FI and SC is given by $\dot{Q}_{\text{FI}} = \dot{E}_{\text{FI}} - \mu_m \dot{N}_m$ and $\dot{Q}_{\text{SC}} = \dot{E}_{\text{SC}} - \mu_s \dot{N}_s$. Here, $\dot{E}_{\text{FI}}$ ($\dot{E}_{\text{SC}}$) is the rate of energy change in FI (SC), $\mu_m$ is the magnon chemical potential in FI, $\mu_{s}$ is the spin accumulation in SC, $\dot{N}_m$ is the rate of magnon number change in FI, and $\dot{N}_s$ is the rate of electron-hole pair change in SC~\cite{Bender2012}. On account of energy and spin conservation, we obtain: $\dot{E}_{\text{SC}} = - \dot{E}_{FI} \equiv \dot{E}$ and $\dot{N}_s = - \dot{N}_{m} \equiv \dot{N} \equiv I_s / \hbar$, with $I_s$ the spin current across the interface. Here, the inclusion of $\mu_{m,s}$ terms in the definition of heat~\cite{Nakata2021} is necessary when considering effects, such as heat generation, up to second order in $\mu_{m,s}$. The physical justification of this contribution, which is well-known for electrons~\cite{Taylor2002,Heikkila2013}, is as follows. A particle added to an ensemble at an energy below the chemical potential needs to absorb energy from other particles in order to respect the statistical distribution enforced by a nonzero chemical potential~\cite{Taylor2002}. Thus, this causes a cooling of the ensemble as a whole. 

The average heat flow from FI to SC is obtained as $\dot{Q} \equiv (\dot{Q}_{\text{SC}} - \dot{Q}_{\text{FI}})/2 = \dot{E} - \bar{\mu} I_s / \hbar$, with $\bar{\mu} \equiv (\mu_m + \mu_s)/2$. Before we evaluate the spin and heat flow across the interface below, we pause to examine the total heat generation $\Delta \dot{Q} \equiv \dot{Q}_{\text{SC}} + \dot{Q}_{\text{FI}} $ in our system. Considering temperature to remain uniform, the heat generation due to chemical potential driven spin transport is simplified to:
\begin{align}
\Delta \dot{Q} & = I_m \Delta \mu = I_m^2 R_m,
\end{align}
where $I_m \equiv I_s / \hbar$, $\Delta \mu \equiv \mu_m - \mu_s$, and the linear response relation $I_m = \Delta \mu / R_m$ will be derived below. This is the spin analogue of Joule heating expression for charge current flow across an interface. However, unlike charge, the spin current across the interface is not conserved in the presence of, e.g., dipolar interaction, which becomes more relevant at low temperatures. In that case, the spin current leaving the FI is not the same as the spin current entering the SC. Hence, we need to amend the spin Joule heating expression to account for this lack of spin current conservation across the FI/SC interface.
	
Since the energy current across the interface is still conserved, as there are no inelastic scattering processes, we obtain:
		\begin{align}
		\Delta \dot{Q} & = \dot{Q}_{\mathrm{FI}} + \dot{Q}_{\mathrm{SC}}, \\
		   & = - \mu_m \dot{N}_m - \mu_s \dot{N}_s, \\
		   & = \mu_m I_m - \mu_s \frac{I_{SC}}{\hbar},
		\end{align}
where we have defined $I_m \equiv - \dot{N}_m$ as the number of magnons disappearing from the FI per unit time, similar to the main text, and $I_{\text{SC}}$ is the spin current entering the SC. In the absence of dipolar interaction, we have spin conservation and $\hbar I_{m} = I_{\text{SC}}$. With dipolar interaction, as detailed in the appendix, we instead obtain the relation $I_{\text{SC}} = \hbar I_m - \delta I_{\text{SC}}$ that results in:
	\begin{align}
	\Delta \dot{Q} & = I_m \Delta \mu + \mu_s \frac{\delta I_\text{SC}}{\hbar}.
	\end{align}
The second term on the right hand side above is a new contribution emerging from the lack of spin conservation across the interface. Consequently, it is also unique to spin Joule heating and is absent in conventional Joule heating owing to conservation of charge currents. 

While for realistic parameters this correction due to the lack of spin current conservation is practically negligible and the spin Joule heating remains positive, it is tempting to imagine a situation where cooling can be achieved by accomplishing a desired large $|\mu_s|$ (using an external mechanism not considered in the present work) that will allow $\Delta \dot{Q}$ to become negative. Such a situation will necessarily entail an external drive to maintain the desired $\mu_s$ and thereby fully respect the laws of thermodynamics. It is yet another interesting feature of spin physics due to a lack of spin conservation under certain conditions. In the following considerations, we disregard the dipolar interactions and the related lack of spin conservation, taking them into account in the more detailed analysis presented in the appendix.

A generalization of this result to an interface between two FIs with different magnon chemical potentials leads to a similar expression for the heating. In the continuum limit, this leads to magnonic spin-Joule heating power per unit volume in the FI bulk:
\begin{align}\label{eq:magJheat}
P & = \pmb{j}_m^2 \rho_m,
\end{align} 
where $\pmb{j}_m = - \pmb{\nabla} \mu_m / \rho_m$ is the magnon current density and $\rho_m$, the magnon spin resistivity.

Equation \eqref{eq:magJheat} is a general result with broad consequences for magnonic spin transport in different materials, hybrids, and regimes. The charge conservation allows for relating electronic Joule heating directly to work done by the external battery~\cite{Heikkila2013}. Such a straightforward identification does not appear possible for magnonic spin transport. Nevertheless, the spin-Joule heating is also derived from the work done by external sources that maintain nonzero chemical potentials in the system. Due to the high sensitivity of superconductor-based thermometers~\cite{Giazotto2006}, the FI/SC hybrids investigated below offer a suitable platform for an experimental measurement of the spin-Joule heating.

\section{Spin and heat currents}
We consider the system shown in \cref{fig:model}(a). A superconductor is placed in contact with a ferromagnetic insulator, and a spin splitting field $h$ is introduced to the former. This system may be influenced by a spin chemical potential on the superconductor side, a nonequilibrium magnon chemical potential on the ferromagnetic insulator side, or a temperature gradient across the system, all of which may result in the flow of heat and spin currents due to exchange interactions between electrons and magnons at the interface. We show that the spin splitting field leads to a significant asymmetry in the transport properties of the heterostructure, with respect to the orientation of $h$. The main physics behind this effect illustrated in \cref{fig:model}(b). Spin can be transmitted between the two materials when, e.g., a spin down quasiparticle in the superconductor experiences a spin flip upon reflection at the interface, accompanied by the creation of a magnon in the ferromagnetic insulator, or vice versa. At low temperatures, this process is suppressed due to the presence of the superconducting gap. However, the size of this gap can be tuned by $h$. In a spin split superconductor, the density of states for the two spin species is shifted relative to each other by a value of $2h$~\cite{Machon2013,Ozaeta2014,Bergeret2018}. This means that the effective gap that must be overcome by a spin-down quasiparticle undergoing a spin flip is increased if the spin splitting field is parallel to the magnetization in the ferromagnetic insulator ($h > 0$), or reduced if it is antiparallel ($h < 0$). Hence, the latter configuration is more amenable to the generation of transport currents. Disregarding dipolar interactions in the main text, we show that their inclusion does not change our key results in the appendix.

We study the interface between the superconductor and the ferromagnetic insulator using a tunneling Hamiltonian approach~\cite{Bender2012,Kamra2016,Ohnuma2017,Kato2019,Kamra2018},
\begin{align}
H = H_{\text{SC}} + H_{\text{FI}} + H_{\text{int}},
\end{align}
where $H_{SC}$ describes a Zeeman split superconductor,
\begin{align}
H_{\text{SC}} =& \sum_{ks}\xi_k c^\dagger_{ks}c_{ks} - \sum_k\left[\Delta c^\dagger_{k\uparrow}c^\dagger_{-k\downarrow} + \Delta^*c_{k\downarrow}c_{-k\uparrow}\right] \nonumber \\
&- \sum_k\sum_{ss'} h\sigma^z_{ss'}c^\dagger_{ks}c_{ks'},
\end{align}
with $\xi_k = \hbar^2\bm{k}^2/2m - \mu$, with chemical potential $\mu$, and $h$ is the exchange field, assumed to be directed along the $z$ axis. The Hamiltonian of the ferromagnetic insulator is given within the Holstein-Primakoff approximation as~\cite{Holstein1940}
\begin{align}
H_{\text{FI}} = \sum_k \hbar\omega_ka_k^\dagger a_k,
\end{align}
with magnon operators $a_k$ and $a_k^\dagger$. We assume a quadratic dispersion of the form $\hbar\omega_k = \Delta_m + J_m\bm{k}^2$ for simplicity. More realistic models for the magnon dispersion, taking, e.g., the dipolar interaction into account (as discussed in the appendix), will modify our results quantitatively, but the qualitative picture presented below remains the same. The two materials may communicate by the exchange of spin, in which a magnon on the ferromagnet side is either absorbed or created by a quasiparticle spin flip on the superconductor side. This process is captured by~\cite{Kato2019} 
\begin{align}
H_{\text{int}} = \sum_{kk'} \left[W_{kk'}s^-_{-k}a_{k'} + W_{kk'}^*s^+_{k}a_{k'}^\dagger\right],
\end{align}
with $s^\pm_k = s^x_k \pm s^y_k$, and $s^j_k = \frac{\hbar}{2}\sum_{qss'}\sigma^j_{ss'} c^\dagger_{ks}c_{k+q,s'}$.

The transport properties of this system are most conveniently studied on the ferromagnetic insulator side, where the tunneling spin current becomes $I_s = i\left\langle\left[N_m,H\right]\right\rangle$, with $N_m =\sum_{k} a^\dagger_k a_k$. This gives,

\begin{align}
I_s &=i\sum_{kk'}\left\langle\left[W_{kk'}s^\dagger_{k}a_{k'} - W_{kk'}^*s_{k}a_{k'}^\dagger\right]\right\rangle.
\end{align}
In a similar way, the average heat current is $\dot{Q} = i\left\langle \left[H_\text{FI},H\right]\right\rangle/\hbar-\bar{\mu}I_s/\hbar$, giving
\begin{align}
\dot{Q} &= \frac{i}{\hbar}\sum_{kk'}\left(\hbar\omega_{k'}-\bar{\mu}\right)\left\langle\left[W_{kk'}s^\dagger_{k}a_{k'} - W_{kk'}^*s_{k}a_{k'}^\dagger\right]\right\rangle.
\end{align}
These quantities are derived in the appendix, resulting in 
\begin{align}
I_s& = 4\pi|W|^2\hbar^3V_{\text{FI}}V_{\text{SC}}^2\nu_0^2\int d\omega\; \nu_m(\hbar\omega)\chi_s(\hbar\omega)\times\nonumber \\
&\hspace{2.5cm}\left[n_{\text{FI}}(\hbar\omega-\mu_m) - n_{\text{SC}}(\hbar\omega - \mu_s)\right],\label{eq:Iz}\\
\dot{Q}& =-4\pi|W|^2\hbar^2 V_{\text{FI}}V_{\text{SC}}^2\nu_0^2\int d\omega\; \left(\hbar\omega-\bar{\mu}\right)\nu_m(\hbar\omega)\chi_s(\hbar\omega)\times \nonumber \\
&\hspace{2.5cm}\left[n_{\text{FI}}(\hbar\omega-\mu_m) - n_{\text{SC}}(\hbar\omega-\mu_s)\right]\label{eq:Q},
\end{align}
under the approximation of $W_{kk'}\simeq W$. Here, $V_{\text{FI}}$ is the volume of the FI, $n_j(\varepsilon) = \left[e^{\varepsilon/k_\text{B}T_j}-1\right]^{-1}$ is the Bose-Einstein distribution function in material $j$, $\nu_m(\varepsilon) =\sqrt{\varepsilon - \Delta_m}/4\pi^2J_m^{3/2}$ is the magnon density of states and
\begin{align}
\chi_s(\hbar\omega) =& \int d\varepsilon\; F_\Delta\nu(\varepsilon+h)\nu(\varepsilon+\hbar\omega-h)\times\nonumber \\
&\left[f(\varepsilon + \hbar\omega - \mu_s/2) - f(\varepsilon + \mu_s/2)\right]
\label{eq:chi}
\end{align}
is (proportional to) the transverse spin susceptibility of the superconductor, with $\nu(\varepsilon) = \Re\left[|\varepsilon|/\sqrt{\varepsilon^2 - |\Delta|^2}\right]$ the superconducting density of states, $F_\Delta = 1 + |\Delta|^2/(\varepsilon +h)(\varepsilon + \hbar\omega - h)$ the coherence factor, and $f(\varepsilon) = \left[e^{\varepsilon/k_{\text{B}}T} + 1\right]^{-1}$ the Fermi-Dirac distribution function.

\begin{figure}[tb]
	\includegraphics[width=\columnwidth]{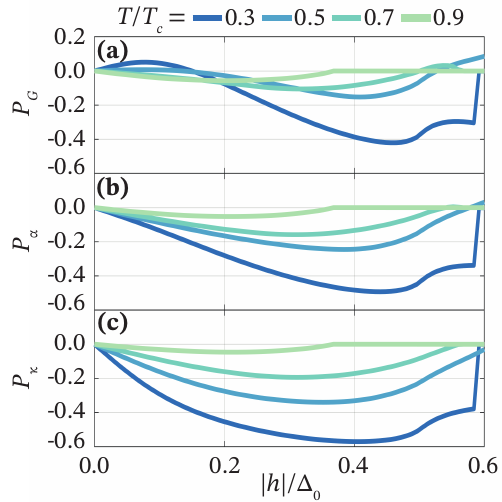}
	\caption{The asymmetry of the linear-response coefficients with respect to the direction of the exchange field $h$, as a function of $|h|$, shown here for a selection of temperatures. Here $P_x = (x_\uparrow - x_\downarrow)/(x_\uparrow + x_\downarrow)$, with the arrows indicating the sign of the exchange field, and $x$ representing (a) the spin conductance $G$, (b) the spin Seebeck coefficient, and (c) the heat conductance. The sharp turns observed in the figure are due to the $h$ crossing the critical value $h_c$ at which superconductivity is destroyed. }
	\label{fig:polar}
\end{figure}

\begin{figure}[tb]
	\includegraphics[width=\columnwidth]{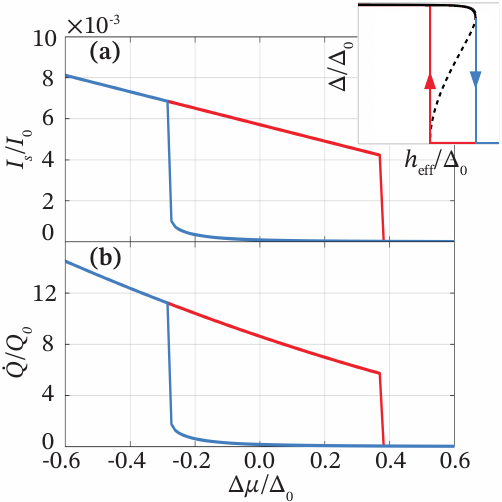}
	\caption{The spin and heat current as a function of the chemical potential difference, $\Delta\mu$, exhibiting a switching effect in which both $I_s$ and $\dot{Q}$ exhibit a jump in magnitude on the order of \si{100}. Here, an exchange field of $h = 0.7\Delta_0$ is applied, which is close to the critical field at which superconductivity is destroyed. We have set $T_{\text{SC}} = 0.1T_c$, $T_{\text{FI}} = 0.5T_c$, and $\mu_m = 0$. The red (blue) curve describes a situation in which superconductivity is regained (destroyed). These two curves are different due to the hysteresis caused by the bistability of the superconducting gap $\Delta$ as a function of the effective spin splitting $h_{\text{eff}} = h - \mu_s/2$, as shown in the inset. Here, $I_0 = \hbar^2|W|^2V_{\text{FI}}V_{\text{SC}}^2\nu_0^2\Delta_0^{5/2}/\pi J_m^{2/3}$, and $Q_0 = I_0\Delta_0/\hbar$.}
	\label{fig:rect}
\end{figure}

For small temperature differences across the tunnel junction, $\Delta T = T_{\text{FI}} - T_{\text{SC}}$, or a small difference in chemical potentials, $\Delta\mu = \mu_m - \mu_s$, \cref{eq:Iz,eq:Q} may be linearized, to obtain
\begin{align}
\begin{pmatrix} I_s \\ \dot{Q}\end{pmatrix} = \begin{pmatrix} G & \alpha \\ -\alpha/\hbar & \kappa T\end{pmatrix}\begin{pmatrix}\Delta\mu \\ \Delta T/T\end{pmatrix}, 
\end{align}
with $T = \left(T_{\text{FI}} + T_{\text{SC}}\right)/2$. This defines the spin conductivity $G$, the spin-dependent Seebeck coefficient $\alpha$, and the heat conductivity $\kappa$ as
\begin{align}
G& = 4\pi\hbar^3|W|^2V_{\text{FI}}V_{\text{SC}}^2\nu_0^2\int d\omega\; \frac{\chi_s(\hbar\omega)\nu_m(\hbar\omega)}{4k_{\text{B}}T\sinh^2\frac{\hbar\omega - \bar{\mu}}{2k_{\text{B}}T}},\label{eq:G} \\
\alpha& = 4\pi\hbar^3|W|^2V_{\text{FI}}V_{\text{SC}}^2\nu_0^2\int d\omega\; \frac{\left(\hbar\omega-\bar{\mu}\right)\chi_s(\hbar\omega)\nu_m(\hbar\omega)}{4k_{\text{B}}T\sinh^2\frac{\hbar\omega - \bar{\mu}}{2k_{\text{B}}T}}, \label{eq:alpha} \\
\kappa& = -4\pi\hbar^2|W|^2V_{\text{FI}}V_{\text{SC}}^2\nu_0^2\int d\omega\; \frac{\left(\hbar\omega-\bar{\mu}\right)^2\chi_s(\hbar\omega)\nu_m(\hbar\omega )}{4k_{\text{B}}T^2\sinh^2\frac{\hbar\omega - \bar{\mu}}{2k_{\text{B}}T}}.\label{eq:kappa}
\end{align}
In the following, we set $\Delta_m = \Delta_0$ for simplicity.

\section{Control over the interfacial conductances and rectification}
To investigate the effect of the spin splitting field $h$ on the transport coefficients given in \cref{eq:G,eq:alpha,eq:kappa}, we define the quantity $P_x = (x_\uparrow - x_\downarrow)/(x_\uparrow + x_\downarrow)$, which we refer to as the polarization of $x$, for $x\in\left\{G,\alpha,\kappa\right\}$, where $\uparrow$ ($\downarrow$) indicates $h > 0$ ($h< 0$). The result is shown in \cref{fig:polar}, for $|h| \in \left[0,0.6\right]$ for a range of temperatures $T$. It is seen that for the lowest temperature considered, $T/T_c = 0.3$, there is a significant polarization. It is negative, indicating that larger currents are to be expected for $h < 0$, consistent with the physical picture presented in \cref{fig:model}(b). We also see that the effect diminishes as the temperature of the heterostructure approaches $T_c$, at which point the superconductor transitions to a normal metal, with no modulation of the density of states, and thus no polarization.

The gap in the density of states, which is present in the superconducting state, but not in the normal metal state, has the potential for an interesting application. In the following, we set the temperature of the superconductor to $T_{\text{SC}} = 0.1T_c$, and in the FI to $T_{\text{FI}} = 0.5T_c$. Hence, a temperature gradient is maintained across the interface, and both spin and heat currents are flowing between the two materials. On the other hand, the magnitude of these currents is largely reduced compared to a normal metal due to the superconducting gap. Next, we set the spin splitting field to $h = 0.7\Delta_0$. This is close to the critical field at which the superconductor transitions to the normal state, but the size of the gap still remains close to the maximal value $\Delta_0$, as evidenced by a self-consistent determination of $\Delta(h)$. We note that the superconducting gap responds to a spin chemical potential as \cite{Bobkova2011} $\Delta(h)\to\Delta(h_{\text{eff}})$, with $h_{\text{eff}} = h - \mu_s/2$. Hence, if $\mu_s < 0$, the two contributions will add up, which has the potential of bringing the superconductor into the normal-state regime at some critical field $h_{\text{eff}}^-$. In the opposite case, $\mu_s$ partially cancels $h$, and the normal-state system returns to the superconducting state at $h_{\text{eff}}^+$. We note that due to the hysteresis caused by the bistability of $\Delta(h_{\text{eff}})$ in the transition region \cite{Sarma1963,Larkin1964,Bobkova2014}, $h_{\text{eff}}^+$ is generally not equal to $h_{\text{eff}}^-$, as indicated in the inset of \cref{fig:rect}. In any case, the point is that $\mu_s$ can cause superconductivity to either be destroyed or regained, which produces an abrupt change in the size of the currents. To illustrate this effect, we plot $I_s$ and $\dot{Q}$, as given by \cref{eq:Iz,eq:Q}, in \cref{fig:rect}(a) and (b), respectively, for $\Delta\mu/\Delta_0\in[-0.6,+0.6]$, keeping $\mu_m = 0$ fixed. We find that both $I_s$ and $\dot{Q}$ feature jumps in magnitude on the order of 100 when a transition takes place, for this parameter set. This abrupt jump in the characteristics correspondingly leads to a huge change of the device linear response conductivities, given by the slope of the characteristic. This is reminiscent of the Zener diode around its peak inverse voltage and embodies the device nonlinearity. Furthermore, the characteristics depicted in \cref{fig:rect} clearly shows the asymmetry with respect to reversing the sign of the drive $\Delta \mu$ resulting in a rectification effect.
To obtain an estimate of the order of magnitude of the rectification effect we consider a typical sample size of $V_{\text{SC}} = V_{\text{FI}} = t\times L_x\times L_y = \SI{10}{\nano\metre}\times\SI{1}{\micro\metre}\times\SI{1}{\micro\metre}$. Using material parameters for aluminium for the superconductor and YIG for the ferromagnetic insulator, we get $\Delta_0\sim \SI{150}{\micro\electronvolt}$, $\nu_0\sim \SI{2e28}{\per\electronvolt\per\metre\cubed}$~\cite{Kittel2005}, and $J_m\sim\SI{8e-40}{\joule\metre\squared}$~\cite{Cornelissen2018}. At room temperature, the spin conductance on Al$|$YIG interfaces has been measured to $Ge^2/\hbar L_yt\sim \SIrange{e12}{e13}{\siemens\per\metre\squared}$~\cite{Dejene2015,Das2019}, and by comparing with the room temperature limit of \cref{eq:G}, $G = 6\pi\zeta\left(\frac{3}{2}\right)\hbar^2W^2\nu_0^2V_{\text{FI}}V_{\text{SC}}^2/\Lambda^3$, with $\Lambda = \sqrt{4\pi J_m/k_{\text{B}}T}$, we obtain a rough estimate of $W\sim\SIrange{e3}{e4}{\per\second}$. This parameter quantifies the interaction between the spins in FI and electrons in S at the interface. It is expected and has been found to be largely temperature independent~\cite{Czeschka2011,Meyer2014}. Thus, it allows us to reliably evaluate quantities at low temperatures considered herein. With this, we estimate the size of the spin and heat currents in \cref{fig:rect} to be on the order of $I_0\simeq\SIrange{1}{10}{\micro\electronvolt}$ and $Q_0\sim\SIrange{0.1}{0.01}{\pico\watt}$. The hundred-fold rectification effect shown in \cref{fig:rect} exploits the smallness of the superconducting gap in accomplishing this feat at a low temperature, assumed $\sim 1$K here. An estimation of the spin and heat currents at higher temperatures, compared to our low temperature estimates discussed here, can be obtained by recognizing:~\cite{Cornelissen2016} $G \sim T^{3/2}$ and $\kappa \sim T^{5/2}$. For example, the estimated spin current at 100 K would then become $I_0 \sim 100^{3/2} (1 - 10) \mu\mathrm{eV} = 1 - 10$ meV. However, our demonstrated rectification works only at temperatures comparable to the superconducting gap. Achieving these good rectifications at high temperature could be accomplished by employing high-Tc superconductors.


\section{Summary} 
Exploiting the weaker energy scale of superconducting gap, we have demonstrated a broad range of nonlinear effects in the context of magnonic spin transport in superconductor-ferromagnetic insulator hybrids. The predicted control over interfacial conductances and hysteretic rectification I-V characteristics open avenues for integrating magnonic devices into unconventional computing architectures, for example. Our theoretical demonstration of magnonic spin-Joule heating provides valuable insights into the wide range of studies and devices involving chemical potential-driven spin transport.

\begin{acknowledgments}
A.K. acknowledges financial support from the Spanish Ministry for Science and Innovation -- AEI Grant CEX2018-000805-M (through the ``Maria de Maeztu'' Programme for Units of Excellence in R\&D). I.V.B. acknowledges financial support from the RSF project No.22-42-04408. Nordita is supported in part by NordForsk.
\end{acknowledgments}

\appendix
\section{Derivation of the spin and heat currents}
Here, we provide details of the calculation of the spin and heat currents. We then proceed to evaluate the linear response and spin Joule heating in the system under consideration. We also include the effect of dipolar interaction. In most previous studies~\cite{Bender2012,Ohnuma2017,Kato2019,Cornelissen2016}, the dipolar interaction has been disregarded. Here, we demonstrate that its inclusion in the model leaves the linear response equations qualitatively unchanged. Hence, the rectification results in the main text remain essentially the same on inclusion of dipolar interactions. More importantly, we find small but novel contributions to spin Joule heating that result because spin current is not a conserved quantity in the presence of dipolar interactions. Such a contribution is precluded in conventional Joule heating since charge, unlike spin, is a conserved quantity.

In the presence of dipolar interactions, the Hamiltonian of the ferromagnetic insulator is modified to~\cite{Kamra2016}
\begin{align}
H_{\text{FI}} = \sum_k A_ka^\dagger_ka_k + B_ka_ka_{-k} + B_k^*a_{k}^\dagger a_{-k}^\dagger,
\end{align}
which can be diagonalized by a Bogoliubov transformation, giving
\begin{align}
H_{\text{FI}} = \sum_k \hbar\Omega_k\alpha_k^\dagger\alpha_k
\end{align}
where $\alpha_k$ is related to the magnon operators via $a_k = u_k\alpha_k + v_k^*\alpha^\dagger_{-k}$, with $u_k = (A_k + \hbar\Omega_k)/\sqrt{(A_k+\hbar\Omega_k)^2 - 4|B_k|^2}$ and $v_k = -2B_k/\sqrt{(A_k+\hbar\Omega_k)^2 - 4|B_k|^2}$. The dispersion of these quasiparticles is furthermore given as $\hbar\Omega_k = \sqrt{A_k^2 - 4|B_k|^2}$.  

In the superconductor, it's convenient to transform the electron operators to a basis which diagonalizes the Hamiltonian, $c_{k\uparrow} = x_k\gamma_{k\uparrow} + y_k\gamma^\dagger_{-k\downarrow}$ and $c_{k\downarrow} = x_k\gamma_{k\downarrow} - y_k\gamma_{-k\uparrow}^\dagger$, with $x_k^2 = \left(1 + \xi_k/E_k\right)/2$ and $y_k^2 = \left(1 - \xi_k/E_k\right)/2$, and $E_k = \sqrt{\xi_k^2 + \Delta^2}$, such that
\begin{align}
H_{\text{SC}} &=\sum_{ks}\left(E_k - sh\right)\gamma^\dagger_{ks}\gamma_{ks} \nonumber\\
&=\frac{1}{2}\sum_{ks}\left(E_k - sh\right)\gamma^\dagger_{ks}\gamma_{ks} - \frac{1}{2}\sum_{ks}\left(E_k - sh\right)\gamma_{ks}\gamma^\dagger_{ks} \nonumber \\
&\equiv \frac{1}{2}\sum_{ks\lambda}\left(\lambda E_k - sh\right)\gamma^\dagger_{ks\lambda}\gamma_{ks\lambda},
\end{align}
where we employ the semiconductor picture, where $\lambda = \pm1$, and $\gamma_{ks+} = \gamma^\dagger_{-k,-s,-}$. For brevity, we derive the equations of motion for $\lambda = +1$ only, and reintroduce the sum over $\lambda$ at the crucial point.

The interaction Hamiltonian is obtained as:
\begin{align}
H_{\text{int}}  &= \hbar \sum_{k_1 k_2 q} W_{k_1 k_2 q} c_{k1,\downarrow}^\dagger c_{k_2,\uparrow} a_q + \mathrm{h.c.}, \\
     = \hbar \sum_{k_1 k_2 q} &\left[W_{k_1 k_2 q} u_q c_{k1,\downarrow}^\dagger c_{k_2,\uparrow} \alpha_q\right. \nonumber \\
&\left. + W_{k_1 k_2 q} v_q^* c_{k1,\downarrow}^\dagger c_{k_2,\uparrow} \alpha_{-q}^\dagger\right] + \mathrm{h.c.}, \\
    = \hbar \sum_{k_1 k_2 q}&\left[ W_{k_1 k_2 q} u_q c_{k1,\downarrow}^\dagger c_{k_2,\uparrow} \alpha_q \right.\nonumber\\ 
&\left.+ W_{k_1 k_2 (-q)}^* v_q c_{k2,\uparrow}^\dagger c_{k_1,\downarrow} \alpha_{q}\right] + \mathrm{h.c.}, \\
= \hbar \sum_{k_1 k_2 q} &\left( W_{k_1 k_2 q} u_q c_{k1,\downarrow}^\dagger c_{k_2,\uparrow} + W_{k_1 k_2 (-q)}^* v_q c_{k2,\uparrow}^\dagger c_{k_1,\downarrow} \right) \alpha_{q} \nonumber \\
&+ \mathrm{h.c.}
\end{align}
where $\mathrm{h.c.}$ stands for hermitian conjugate. Here, we have employed the property: $v_{-q} = v_{q}$. The interaction Hamiltonian can be expressed in terms of the eigenmode ladder operators as:
\begin{align}
H_{\text{int}} = \hbar \sum_{k_1 k_2 q} \alpha_{q} Z + \mathrm{h.c.},
\end{align}
where $Z = Z_1 + Z_2 + Z_3 + Z_4$ with
\begin{align}
Z_1 & = W_{k_1 k_2 q} u_q \left( x^*_{k_1} x_{k_2} \gamma^\dagger_{k_1,\downarrow} \gamma_{k_2,\uparrow} - y_{k_2} y^*_{k_1} \gamma_{-k_1,\uparrow} \gamma^\dagger_{-k_2,\downarrow} \right) \nonumber \\
&\equiv u_q M_{k_1k_2q}\gamma^\dagger_{k_1,\downarrow}\gamma_{k_2,\uparrow}, \\
Z_2 & = W_{k_1 k_2 (-q)}^* v_q \left( x^*_{k_2} x_{k_1} \gamma^\dagger_{k_2,\uparrow} \gamma_{k_1,\downarrow} - y_{k_2}^* y_{k_1} \gamma_{-k_2,\downarrow} \gamma^\dagger_{-k_1,\uparrow} \right)\nonumber \\
&\equiv v_qM^*_{k_1k_2(-q)}\gamma^\dagger_{k_2,\uparrow}\gamma_{k_1,\downarrow}, \\
Z_3 & = W_{k_1 k_2 q} u_q  x^*_{k_1} y_{k_2} \gamma^\dagger_{k_1,\downarrow} \gamma^\dagger_{-k_2,\downarrow} \nonumber \\
&- W_{k_1 k_2 (-q)}^* v_q  x^*_{k_2} y_{k_1} \gamma^\dagger_{k_2,\uparrow} \gamma^\dagger_{-k_1,\uparrow}, \\
Z_4 & = - W_{k_1 k_2 q} u_q  x_{k_2} y_{k_1}^* \gamma_{-k_1,\uparrow} \gamma_{k_2,\uparrow} \nonumber \\
&+ W_{k_1 k_2 (-q)}^* v_q  y^*_{k_2} x_{k_1}  \gamma_{-k_2,\downarrow}\gamma_{k_1,\downarrow},
\end{align}
where $M_{k_1k_2q} = W_{k_1k_2q}x_{k_1}^*x_{k_2} + W_{-k_2(-k_1)q}y_{k_1}y_{k_2}^*$.

The operator for spin current injected by the FI into the superconductor is given by
\begin{align}
\hat{I}_{\text{SC}} &= \dot{\mathcal{S}}_z = \frac{1}{i\hbar} \left[ \mathcal{S}_z , H_{\mathrm{int}} \right],
\end{align}
where $\mathcal{S}_z \equiv (\hbar/2) \sum_{\pmb{k}} c^\dagger_{k\uparrow} c_{k\uparrow} - c^\dagger_{k\downarrow}c_{k\downarrow}$. In order to evaluate the spin current, we need to take the expectation value of the spin current operator, $I_{\text{SC}} = \left\langle\hat{I}_{\text{SC}}\right\rangle$, thus obtained:

\begin{align}
I_{\text{SC}} &= -\frac{i}{2}\sum_k\left\langle \left[c^\dagger_{k\uparrow} c_{k\uparrow} - c^\dagger_{k\downarrow}c_{k\downarrow}\;,\;H_{\mathrm{int}} \right]\right\rangle \nonumber \\
=i\hbar\sum_{k_1k_2q}&\left(W_{k_2k_1q}u_{q}\left\langle c^\dagger_{k_1,\downarrow}c_{k_2,\uparrow}\alpha_q\right\rangle\right.\nonumber \\
&\left. - W_{k_1k_2(-q)}^*v_q\left\langle c^\dagger_{k_2,\uparrow}c_{k_1,\downarrow}\alpha_{q}\right\rangle\right) -\mathrm{h.c.}\nonumber \\
=-2\hbar\sum_{k_1k_2q}&\Im\left[\left\langle W_{k_2k_1q}u_{q}c^\dagger_{k_1,\downarrow}c_{k_2,\uparrow}\alpha_q\right\rangle\right. \nonumber \\
&\left.-W_{k_1k_2(-q)}^*v_q\left\langle c^\dagger_{k_2,\uparrow}c_{k_1,\downarrow}\alpha_{q}\right\rangle\right]
\label{eq:Isdip}
\end{align}
In terms of the eigenmode ladder operators of the superconductor, this becomes
\begin{align}
I_{\text{SC}} &= -2\hbar\sum_{k_1k_2q}\Im\left[\left\langle\alpha_q\left(Z_1 - Z_2 + \bar{Z}_3 - \bar{Z}_4\right)\right\rangle\right]\nonumber \\
&\equiv I_{\text{SC}}^{(1)} +I_{\text{SC}}^{(2)} +I_{\text{SC}}^{(3)} +I_{\text{SC}}^{(4)} ,
\end{align}
where
\begin{align}
\bar{Z}_3 & = W_{k_1 k_2 q} u_q  x^*_{k_1} y_{k_2} \gamma^\dagger_{k_1,\downarrow} \gamma^\dagger_{-k_2,\downarrow} \nonumber \\
&+ W_{k_1 k_2 (-q)}^* v_q  x^*_{k_2} y_{k_1} \gamma^\dagger_{k_2,\uparrow} \gamma^\dagger_{-k_1,\uparrow}, \\
\bar{Z}_4 & = W_{k_1 k_2 q} u_q  x_{k_2} y_{k_1}^* \gamma_{-k_1,\uparrow} \gamma_{k_2,\uparrow} \nonumber \\
&+ W_{k_1 k_2 (-q)}^* v_q  y^*_{k_2} x_{k_1}  \gamma_{-k_2,\downarrow}\gamma_{k_1,\downarrow}. 
\end{align}
On the ferromagnetic insulator side, due to the spin-nonconserving terms arising due to the dipolar interaction, the spin current traversing the interface is not the same as what is finally flows in the FI. We know that carrying out the expectation value has to be done in the eigenbasis of the FI. Thus, the appropriate operator for the spin current injected into the FI is obtained as:

\begin{align}
\hat{I}_{\text{FI}} &= -\hbar\sum_{\pmb{k}} (1 + 2 |v_{\pmb{k}}|^2) \dot{N_{k}} = \sum_{\pmb{k}}  i (1 + 2 |v_{\pmb{k}}|^2)  \left[ N_{k} , H_{\mathrm{int}} \right],
\end{align}
where $N_{k} \equiv \alpha_k^\dagger \alpha_k$. Hence, its expectation value becomes
\begin{align}
I_{\text{FI}} &= -i\hbar\sum_{k_1k_2q}\left\langle\left(1 + 2|v_q|^2\right)\alpha_qZ - \mathrm{h.c.}\right\rangle \\ 
& = 2\hbar\sum_{k_1k_2q}\Im\left[\left\langle\left(1 + 2|v_q|^2\right)\alpha_qZ\right\rangle\right] \nonumber \\
&\equiv I_{\text{FI}}^{(1)} + I_{\text{FI}}^{(2)} + I_{\text{FI}}^{(3)} + I_{\text{FI}}^{(4)}.
\end{align}

The time evolution of $\gamma_{k\uparrow}$ is found from the Heisenberg equation as
\begin{align}
&i\hbar\partial_t\gamma_{k\uparrow} = \left[\gamma_{k\uparrow}\;,\;H\right] = E_{k,\uparrow}\gamma_{k,\uparrow} \nonumber \\
&+ \hbar\sum_{k_1q}\left[u_q M^*_{k_1kq}\alpha_q^\dagger + v_qM^*_{k_1k(-q)}\alpha_q\right]\gamma_{k_1,\downarrow}\nonumber \\
&-\hbar\sum_{k_1q}\left[ u_qN^*_{k_1kq}\alpha^\dagger_q + v_qN^*_{k_1k(-q)}\alpha_q\right]\gamma_{-k_1,\uparrow}^\dagger,
\end{align}
with $E_{k,\uparrow} = E_k - h$ and $N_{k_1k_2q} = W_{k_1k_2q}x_{k_1}^*y_{k_2} - W_{-k_2(-k_1)q}x_{k_2}^*y_{k_1}$. Its solution is given as
\begin{align}
&\gamma_{k,\uparrow}(t) = e^{-iE_{k,\uparrow}(t-t_0)/\hbar}\gamma_{k,\uparrow}(t_0) \nonumber \\
&-i\sum_{k_1q}\int_{t_0}^t dt'\;e^{-iE_{k,\uparrow}(t-t')/\hbar}\left[u_q M^*_{k_1kq}\alpha_q^\dagger(t') \right. \nonumber \\
&\left.+ v_qM^*_{k_1k(-q)}\alpha_q(t')\right]\gamma_{k_1,\downarrow}(t') \nonumber\\
&+
\left[ u_qN^*_{k_1kq}\alpha^\dagger_q(t') + v_qN^*_{k_1k(-q)}\alpha_q(t')\right]\gamma_{-k_1,\uparrow}^\dagger(t').
\end{align}
\begin{widetext}

To lowest order in $W_{k_1k_2q}$ we obtain
\begin{align}
\gamma_{k,\uparrow}(t) =& e^{-iE_{k,\uparrow}(t-t_0)/\hbar}\left[\gamma_{k,\uparrow}(t_0) - i\pi\sum_{k_1q}u_qM_{k_1kq}^*\alpha_q^\dagger(t_0)\gamma_{k_1,\downarrow}(t_0)\delta\left(\frac{E_{k,\uparrow}-E_{k_1,\downarrow}}{\hbar} + \Omega_q\right)\right. \nonumber \\
&-i\pi\sum_{k_1q}v_qM_{k_1k(-q)}^*\alpha_q(t_0)\gamma_{k1,\downarrow}(t_0)\delta\left(\frac{E_{k,\uparrow}-E_{k_1,\downarrow}}{\hbar} - \Omega_q\right) \nonumber \\
&+i\pi\sum_{k_1q}u_qN_{kk_1q}^*\alpha^\dagger_q(t_0)\gamma_{-k1,\uparrow}^\dagger(t_0)\delta\left(\frac{E_{k,\uparrow}+E_{k_1,\uparrow}}{\hbar} + \Omega_q\right) \nonumber \\
&\left.+i\pi\sum_{k_1q}v_qN_{kk_1(-q)}^*\alpha_q(t_0)\gamma_{-k1,\uparrow}^\dagger(t_0)\delta\left(\frac{E_{k,\uparrow}+E_{k_1,\uparrow}}{\hbar} - \Omega_q\right)\right].
\end{align}
Similarly, $\gamma_{k,\downarrow}$ is found to be
\begin{align}
\gamma_{k,\downarrow}(t) =& e^{-iE_{k,\downarrow}(t-t_0)/\hbar}\left[\gamma_{k,\downarrow}(t_0) - i\pi\sum_{k_1q}u_qM_{kk_1q}\alpha_q(t_0)\gamma_{k_1,\uparrow}(t_0)\delta\left(\frac{E_{k,\downarrow}-E_{k_1,\uparrow}}{\hbar} - \Omega_q\right)\right. \nonumber \\
&-i\pi\sum_{k_1q}v_q^*M_{kk_1(-q)}\alpha_q^\dagger(t_0)\gamma_{k1,\uparrow}(t_0)\delta\left(\frac{E_{k,\downarrow}-E_{k_1,\uparrow}}{\hbar} + \Omega_q\right) \nonumber \\
&-i\pi\sum_{k_1q}u_qN_{kk_1q}\alpha_q(t_0)\gamma_{-k1,\downarrow}^\dagger(t_0)\delta\left(\frac{E_{k,\downarrow}+E_{k_1,\downarrow}}{\hbar} - \Omega_q\right) \nonumber \\
&\left.-i\pi\sum_{k_1q}v_q^*N_{kk_1(-q)}\alpha_q^\dagger(t_0)\gamma_{-k1,\downarrow}^\dagger(t_0)\delta\left(\frac{E_{k,\downarrow}+E_{k_1,\downarrow}}{\hbar} + \Omega_q\right)\right].
\end{align}
Hence, the time evolution of $Z_1$ becomes
\begin{align}
Z_1(t) =& -i\pi e^{-i(E_{k_2,\uparrow}-E_{k_1,\downarrow})(t-t_0)/\hbar}\sum_{q_1}\left[u_qu_{q_1}M_{k_1k_2q}M^*_{k_1k_2q_1}\alpha_{q_1}^\dagger(t_0)\gamma_{k_1,\downarrow}^\dagger(t_0)\gamma_{k_1,\downarrow}(t_0)\delta\left(\frac{E_{k_2,\uparrow} - E_{k_1,\downarrow}}{\hbar} + \Omega_{q_1}\right)\right. \nonumber \\
&+u_qv_{q_1}M_{k_1k_2q}M^*_{k_1k_2(-q_1)}\alpha_{q_1}(t_0)\gamma_{k_1,\downarrow}^\dagger(t_0)\gamma_{k_1,\downarrow}(t_0)\delta\left(\frac{E_{k_2,\uparrow} - E_{k_1,\downarrow}}{\hbar} - \Omega_{q_1}\right) \nonumber \\
&-u_qu_{q_1}M_{k_1k_2q}M^*_{k_1k_2q_1}\alpha_{q_1}^\dagger(t_0)\gamma_{k_2,\uparrow}^\dagger(t_0)\gamma_{k_2,\uparrow}(t_0)\delta\left(\frac{E_{k_1,\downarrow} - E_{k_2,\uparrow}}{\hbar} - \Omega_{q_1}\right) \nonumber \\
&-u_qv_{q_1}M_{k_1k_2q}M^*_{k_1k_2(-q_1)}\alpha_{q_1}(t_0)\gamma_{k_2,\uparrow}^\dagger(t_0)\gamma_{k_2,\uparrow}(t_0)\delta\left(\frac{E_{k_1,\downarrow} - E_{k_2,\uparrow}}{\hbar} + \Omega_{q_1}\right),
\end{align}
\end{widetext}
where we have anticipated that only the averages of $\gamma^\dagger_{k,\sigma}(t_0)\gamma_{k,\sigma}(t_0)$ will survive. In a similar vein, with
\begin{align}
&\alpha_q(t) = e^{-i\Omega_q(t-t_0)}\left[\alpha_q(t_0)\right.\nonumber\\
& - i\pi\sum_{k_1k_2}u_qM^*_{k_1k_2q}\gamma^\dagger_{k_2\uparrow}\gamma_{k_1\downarrow}\delta\left(\frac{E_{k_2,\uparrow} - E_{k_1,\downarrow}}{\hbar} + \Omega_q\right) \nonumber \\
&+v_q^*M_{k_1k_2(-q)}\gamma_{k_1,\downarrow}^\dagger\gamma_{k_2,\uparrow}\delta\left(\frac{E_{k_2,\uparrow} - E_{k_1,\downarrow}}{\hbar} - \Omega_q\right) \nonumber \\
&+u_qW^*_{k_1k_2q}x_{k_1}y_{k_2}^*\gamma_{-k_2\downarrow}\gamma_{k_1\downarrow}\delta\left(\frac{E_{k_1,\downarrow} + E_{k_2,\downarrow}}{\hbar} - \Omega_q\right) \nonumber \\
&-v_q^*W_{k_1k_2q}x_{k_2}y_{k_1}^*\gamma_{-k_1\uparrow}\gamma_{k_2\uparrow}\delta\left(\frac{E_{k_2,\uparrow} + E_{k_2,\uparrow}}{\hbar} - \Omega_q\right) \nonumber \\
&-u_qW^*_{k_1k_2q}x^*_{k_2}y_{k_1}\gamma_{k_2\uparrow}^\dagger\gamma_{-k_1\uparrow}^\dagger\delta\left(\frac{E_{k_2,\uparrow} + E_{k_1,\uparrow}}{\hbar} + \Omega_q\right) \nonumber \\
&+v_q^*W_{k_1k_2q}x_{k_1}^*y_{k_2}\gamma_{k_1\downarrow}^\dagger\gamma_{-k_2\downarrow}^\dagger\delta\left(\frac{E_{k_2,\downarrow} + E_{k_1,\downarrow}}{\hbar} + \Omega_q\right) \nonumber \\
\end{align}
we can compute
\begin{align}
&Z_1(t)\alpha_q(t) = -i\pi u_q^2|M_{k_1k_2q}|^2\times\nonumber\\
&\left[\alpha_q^\dagger(t_0)\alpha_q(t_0)\left(\gamma_{k_1,\downarrow}^\dagger(t_0)\gamma_{k_1,\downarrow}(t_0) - \gamma_{k_2,\uparrow}^\dagger(t_0)\gamma_{k_2,\uparrow}(t_0)\right)\right. \nonumber\\
&\left.+\gamma^\dagger_{k_1,\downarrow}\gamma_{k_2,\uparrow}\gamma_{k_2,\uparrow}^\dagger\gamma_{k_1,\downarrow}\right]\delta\left(\frac{E_{k_2,\uparrow} - E_{k_1,\downarrow}}{\hbar} + \Omega_{q_1}\right).
\end{align}
By a similar analysis one finds
\begin{align}
&Z_2(t)\alpha_q(t) = +i\pi|v_q|^2|M_{k_1k_2q}|^2\times\nonumber\\
&\left[\alpha_q^\dagger(t_0)\alpha_q(t_0)\left(\gamma_{k_1,\downarrow}^\dagger(t_0)\gamma_{k_1,\downarrow}(t_0) - \gamma_{k_2,\uparrow}^\dagger(t_0)\gamma_{k_2,\uparrow}(t_0)\right)\right.\nonumber\\
&\left.-\gamma^\dagger_{k_2,\uparrow}\gamma_{k_1,\downarrow}\gamma_{k_1,\downarrow}^\dagger\gamma_{k_2,\uparrow}\right]\delta\left(\frac{E_{k_2,\uparrow} - E_{k_1,\downarrow}}{\hbar} - \Omega_{q_1}\right).
\end{align}
In the following, we introduce the approximation $W_{k_1k_2q} = W$, in which case we find
\begin{align*}
|M_{k_1k_2q}|^2 &= |W|^2\left[\frac{1}{2} + \frac{\xi_{k_1}\xi_{k_2} + |\Delta|^2}{2E_{k_1}E_{k_2}}\right]\equiv |W|^2 F_{k_1k_2}, \\
|N_{k_1k_2q}|^2 &= |W|^2\left[\frac{1}{2} - \frac{\xi_{k_1}\xi_{k_2} + |\Delta|^2}{2E_{k_1}E_{k_2}}\right]\equiv |W|^2 G_{k_1k_2}.
\end{align*}
Next, we use
\begin{align}
&\left\langle \gamma_{k,\sigma}^\dagger(t_0)\gamma_{k,\sigma}(t_0)\right\rangle = f(E_{k\sigma} - \mu_\sigma) \\
&\left\langle \alpha_q^\dagger(t_0) \alpha_q\right\rangle = n(\hbar\Omega_q - \mu_m) \\
&\left\langle \gamma^\dagger_{k_1,\downarrow}\gamma_{k_2,\uparrow}\gamma_{k_2,\uparrow}^\dagger\gamma_{k_1,\downarrow}\right\rangle = -\left[f(E_{k_1\downarrow} - \mu_\downarrow) - f(E_{k_2\uparrow} - \mu_\uparrow)\right]\nonumber\\
&\times n(+E_{k_1\downarrow}-E_{k_2\uparrow} - \mu_\downarrow + \mu_\uparrow) \\
&\left\langle \gamma^\dagger_{k_2,\uparrow}\gamma_{k_1,\downarrow}\gamma_{k_1,\downarrow}^\dagger\gamma_{k_2,\uparrow}\right\rangle = +\left[f(E_{k_1\downarrow} - \mu_\downarrow) - f(E_{k_2\uparrow} - \mu_\uparrow)\right]\nonumber\\
&\times n(-E_{k_1\downarrow}+E_{k_2\uparrow} + \mu_\downarrow - \mu_\uparrow)
\end{align}
with $f(x) = \left(e^{\beta x} + 1\right)^{-1}$ and $n(x) =\left( e^{\beta x} - 1\right)^{-1}$, to get 
\begin{align}
I_{\text{SC}}^{(1)} = 2\pi\hbar^2& |W|^2\sum_{k_1k_2q} u_q^2F_{k_1k_2}\nonumber\\
&\times\left[f(E_{k_1\downarrow} - \mu_\downarrow) - f(E_{k_2\uparrow} - \mu_\uparrow)\right]\nonumber\\
&\times\left( n_{\text{FI}}(\hbar\Omega_q - \mu_m) - n_{\text{SC}}(\hbar\Omega_q - \mu_s)\right)\nonumber\\
&\times\delta\left(E_{k_1\downarrow} - E_{k_2\uparrow} - \hbar\Omega_q\right) \\ 
I_{\text{SC}}^{(2)} = 2\pi\hbar^2& |W|^2\sum_{k_1k_2q} |v_q|^2F_{k_1k_2}\nonumber\\
&\times\left[f(E_{k_1\downarrow} - \mu_\downarrow) - f(E_{k_2\uparrow} - \mu_\uparrow)\right]\nonumber\\
&\times\left( n_{\text{FI}}(\hbar\Omega_q - \mu_m) - n_{\text{SC}}(\hbar\Omega_q + \mu_s)\right)\nonumber\\
&\times\delta\left(E_{k_1\downarrow} - E_{k_2\uparrow} + \hbar\Omega_q\right) 
\end{align}
and
\begin{align}
I_{\text{FI}}^{(1)} = -2\pi\hbar^2 &|W|^2\sum_{k_1k_2q} \left(1 + 2|v_q|^2\right)u_q^2F_{k_1k_2}\nonumber\\
&\times\left[f(E_{k_1\downarrow} - \mu_\downarrow) - f(E_{k_2\uparrow} - \mu_\uparrow)\right]\nonumber\\
&\times\left( n_{\text{FI}}(\hbar\Omega_q - \mu_m) - n_{\text{SC}}(\hbar\Omega_q - \mu_s)\right) \nonumber \\
&\times\delta\left(E_{k_1\downarrow} - E_{k_2\uparrow} - \hbar\Omega_q\right) \\ 
I_{\text{FI}}^{(2)} = 2\pi\hbar^2 &|W|^2\sum_{k_1k_2q} \left(1 + 2|v_q|^2\right)|v_q|^2F_{k_1k_2}\nonumber\\
&\times\left[f(E_{k_1\downarrow} - \mu_\downarrow) - f(E_{k_2\uparrow} - \mu_\uparrow)\right]\nonumber\\
&\left( n_{\text{FI}}(\hbar\Omega_q - \mu_m) - n_{\text{SC}}(\hbar\Omega_q + \mu_s)\right) \nonumber \\
&\times\delta\left(E_{k_1\downarrow} - E_{k_2\uparrow} + \hbar\Omega_q\right)
\end{align}
For the anomalous part of the currents we get
\begin{align}
I_{\text{SC}}^{(3)} = \pi\hbar^2& |W|^2\sum_{k_1k_2q}u_q^2G_{k_1k_2}\nonumber\\
&\times\left[f(E_{k_1\downarrow}-\mu_\downarrow) - f(-E_{k_2\downarrow}+\mu_\downarrow)\right]\nonumber\\
&\times\left(n_{\text{FI}}(\hbar\Omega - \mu_m) - n_{\text{SC}}(\hbar\Omega - \mu_s)\right)\nonumber\\
&\times\delta\left(E_{k_1\downarrow} + E_{k_2\downarrow} - \hbar\Omega\right) \nonumber \\
-\pi\hbar^2& |W|^2\sum_{k_1k_2q}|v_q|^2G_{k_1k_2}\nonumber\\
&\times\left[f(E_{k_1\uparrow}-\mu_\uparrow) - f(-E_{k_2\uparrow}+\mu_\uparrow)\right]\nonumber\\
&\times\left(n_{\text{FI}}(\hbar\Omega - \mu_m) - n_{\text{SC}}(\hbar\Omega + \mu_s)\right)\nonumber\\
&\times\delta\left(E_{k_1\uparrow} + E_{k_2\uparrow} - \hbar\Omega\right) \\
I_{\text{SC}}^{(4)} = \pi\hbar^2& |W|^2\sum_{k_1k_2q}u_q^2G_{k_1k_2}\nonumber\\
&\times\left[f(E_{k_1\uparrow}-\mu_\uparrow) - f(-E_{k_2\uparrow}+\mu_\uparrow)\right]\nonumber\\
&\times\left(n_{\text{FI}}(\hbar\Omega - \mu_m) - n_{\text{SC}}(\hbar\Omega - \mu_s)\right)\nonumber\\
&\times\delta\left(E_{k_1\uparrow} + E_{k_2\uparrow} + \hbar\Omega\right) \nonumber \\
-\pi\hbar^2& |W|^2\sum_{k_1k_2q}|v_q|^2G_{k_1k_2}\nonumber\\
&\times\left[f(E_{k_1\downarrow}-\mu_\downarrow) - f(-E_{k_2\downarrow}+\mu_\downarrow)\right]\nonumber\\
&\times\left(n_{\text{FI}}(\hbar\Omega - \mu_m) - n_{\text{SC}}(\hbar\Omega + \mu_s)\right)\nonumber\\
&\times\delta\left(E_{k_1\downarrow} + E_{k_2\downarrow} + \hbar\Omega\right)
\end{align}
and
\begin{align}
I_{\text{FI}}^{(3)} = -\pi\hbar^2& |W|^2\sum_{k_1k_2q}\left(1 + 2|v_q|^2\right)u_q^2G_{k_1k_2}\nonumber\\
&\times\left[f(E_{k_1\downarrow}-\mu_\downarrow) - f(-E_{k_2\downarrow}+\mu_\downarrow)\right]\nonumber\\
&\times\left(n_{\text{FI}}(\hbar\Omega - \mu_m) - n_{\text{SC}}(\hbar\Omega - \mu_s)\right)\nonumber\\
&\times\delta\left(E_{k_1\downarrow} + E_{k_2\downarrow} - \hbar\Omega\right) \nonumber \\
-\pi\hbar^2 &|W|^2\sum_{k_1k_2q}\left(1 + 2|v_q|^2\right)|v_q|^2G_{k_1k_2}\nonumber\\
&\times\left[f(E_{k_1\uparrow}-\mu_\uparrow) - f(-E_{k_2\uparrow}+\mu_\uparrow)\right]\nonumber\\
&\left(n_{\text{FI}}(\hbar\Omega - \mu_m) - n_{\text{SC}}(\hbar\Omega + \mu_s)\right)\nonumber\\
&\times\delta\left(E_{k_1\uparrow} + E_{k_2\uparrow} - \hbar\Omega\right) \\
I_{\text{FI}}^{(4)} = \pi\hbar^2& |W|^2\sum_{k_1k_2q}\left(1 + 2|v_q|^2\right)u_q^2G_{k_1k_2}\nonumber\\
&\times\left[f(E_{k_1\uparrow}-\mu_\uparrow) - f(-E_{k_2\uparrow}+\mu_\uparrow)\right]\nonumber\\
&\times\left(n_{\text{FI}}(\hbar\Omega - \mu_m) - n_{\text{SC}}(\hbar\Omega - \mu_s)\right)\nonumber\\
&\times\delta\left(E_{k_1\uparrow} + E_{k_2\uparrow} + \hbar\Omega\right) \nonumber \\
+\pi\hbar^2 &|W|^2\sum_{k_1k_2q}\left(1 + 2|v_q|^2\right)|v_q|^2G_{k_1k_2}\nonumber\\
&\times\left[f(E_{k_1\downarrow}-\mu_\downarrow) - f(-E_{k_2\downarrow}+\mu_\downarrow)\right]\nonumber\\
&\times\left(n_{\text{FI}}(\hbar\Omega - \mu_m) - n_{\text{SC}}(\hbar\Omega + \mu_s)\right)\nonumber\\
&\times\delta\left(E_{k_1\downarrow} + E_{k_2\downarrow} + \hbar\Omega\right)
\end{align}

The next step is to convert the momentum sums into energy integrals, and we use $\sum_k \to \nu_0V_{\text{SC}}\int_{-\infty}^\infty d\xi_k$, where $\nu_0 = mk_{\text{F}}/2\pi^2\hbar^2$ is the density of states at the Fermi level. Furthermore, to produce the correct expressions for the current in the semiconductor picture, we make the substitutions $E_{ki}\to\lambda_i E_{ki}$ for $i\in\{1,2\}$, and sum over $\lambda_1$ and $\lambda_2$. For $I_{\text{SC}}^{(1)}$ we obtain
\begin{align}
I_{\text{SC}}^{(1)} &= 2\pi\hbar^2|W|^2\nu_0^2V_{\text{SC}}^2\sum_{q\lambda}u_q^2\nonumber\\
&\times\int_{-\infty}^\infty d\xi_k\;\Re\left[\frac{|\lambda E_k+\hbar\Omega_q - 2h|}{\sqrt{(\lambda E_k + \hbar\Omega_q - 2h)^2 - \Delta^2}}\right]\nonumber\\
&\times\left(1 + \frac{\Delta^2}{\lambda E_k(\lambda E_k + \hbar\Omega_q - 2h)}\right) \nonumber \\
&\times \left[f(\lambda E_k + \hbar\Omega_q - h - \mu_\downarrow) - f(\lambda E_k - h - \mu_\uparrow)\right]\nonumber\\
&\times\left( n_{\text{FI}}(\hbar\Omega_q - \mu_m) - n_{\text{SC}}(\hbar\Omega_q - \mu_s)\right) \nonumber \\
&\equiv 4\pi\hbar^2|W|^2\nu_0^2V_{\text{SC}}^2\sum_qu_q^2\nonumber\\
&\times\int_{-\infty}^\infty dE_k\;\nu( E_k+h)\nu(E_k + \hbar\Omega_q -h)F(E_k,+h) \nonumber \\
&\times \left[f(E_k + \hbar\Omega_q - \mu_\downarrow) - f(E_k - \mu_\uparrow)\right]\nonumber\\
&\times\left( n_{\text{FI}}(\hbar\Omega_q - \mu_m) - n_{\text{SC}}(\hbar\Omega_q - \mu_s)\right),
\label{eq:isc1}
\end{align}
where we have used that \begin{align*}\sum_\lambda\int_{-\infty}^\infty d\xi_k f(\lambda E_k) &= 2\sum_\lambda\int_0^\infty dE_k \nu(E_k) f(\lambda E_k) \nonumber \\
&= 2\int_{-\infty}^\infty dE_k \nu(E_k) f(E_k,+h),\end{align*} 
with $E_k = \sqrt{\xi_k^2 + \Delta^2}$, $\nu(x) = \Re\left(|x|/\sqrt{x^2 - \Delta^2}\right)$, and $F(x,\pm h) = 1 + \Delta^2/(E_k\pm h)(x \mp h)$. In the same way we get
\begin{align}
I_{\text{SC}}^{(2)} &= -4\pi\hbar^2 |W|^2\nu_0^2V_{\text{SC}}^2\sum_{q} |v_q|^2\nonumber\\
&\times\int_{-\infty}^\infty \nu(E_k-h)\nu(E_k + \hbar\Omega_q + h)F(E_k,-h)\nonumber \\
&\times\left[f(E_k + \hbar\Omega_q + \mu_\downarrow) - f(E_k + \mu_\uparrow)\right]\nonumber\\
&\times\left( n_{\text{FI}}(\hbar\Omega_q - \mu_m) - n_{\text{SC}}(\hbar\Omega_q + \mu_s)\right)
\label{eq:isc2}
\end{align}
Consider next the part of $I_{\text{SC}}^{(3)}$ containing to $u_q^2$. Here, we get
\begin{align}
I_{\text{SC},u}^{(3)}&= 2\pi\hbar^2|W|^2\nu_0^2V_{\text{SC}}^2\sum_qu_q^2\int_{-\infty}^\infty dE_k\;\nonumber\\
&\nu(E_k)\nu(-E_k + \hbar\Omega_q - 2h)G(-E_k + \hbar\Omega_q - 2h) \nonumber \\
&\times \left[f(-E_k + \hbar\Omega_q - h - \mu_\downarrow) - f(-E_k - h + \mu_\downarrow)\right]\nonumber\\
&\times\left( n_{\text{FI}}(\hbar\Omega_q - \mu_m) - n_{\text{SC}}(\hbar\Omega_q - \mu_s)\right),
\end{align}
with $G(x) = 1 - \Delta^2/ E_kx$. Similarly,
\begin{align}
I_{\text{SC},u}^{(4)}&= 2\pi\hbar^2|W|^2\nu_0^2V_{\text{SC}}^2\sum_qu_q^2\int_{-\infty}^\infty dE_k\;\nonumber\\
&\nu(E_k)\nu(-E_k - \hbar\Omega_q + 2h)G(-E_k - \hbar\Omega_q + 2h) \nonumber \\
&\times \left[f(-E_k - \hbar\Omega_q + h - \mu_\uparrow) - f(-E_k + h + \mu_\uparrow)\right]\nonumber\\
&\times\left( n_{\text{FI}}(\hbar\Omega_q - \mu_m) - n_{\text{SC}}(\hbar\Omega_q - \mu_s)\right).
\end{align}
Note that by using the identity $\mu_\uparrow = -\mu_\downarrow$, it is seen that $I_{\text{SC},u}^{(4)} = -I_{\text{SC},u}^{(3)}$. Hence, these two terms cancel. The same analysis gives similarly $I_{\text{SC},v}^{(4)} = -I_{\text{SC},v}^{(3)}$. Repeating the above procedure on the FI-side reveals that $I_{\text{FI}}^{(3)}$ and $I_{\text{FI}}^{(4)}$ also cancel. Moreover, the epxressions for $I_{\text{FI}}^{(1)}$ and $I_{\text{FI}}^{(2)}$ are given as
\begin{align}
I_{\text{FI}}^{(1)}&= -4\pi\hbar^2|W|^2\nu_0^2V_{\text{SC}}^2\sum_q(1+2|v_q|^2)u_q^2\nonumber\\
&\times\int_{-\infty}^\infty dE_k\;\nu( E_k+h)\nu(E_k + \hbar\Omega_q -h)F(E_k,+h) \nonumber \\
&\times \left[f(E_k + \hbar\Omega_q - \mu_\downarrow) - f(E_k - \mu_\uparrow)\right]\nonumber\\
&\times\left( n_{\text{FI}}(\hbar\Omega_q - \mu_m) - n_{\text{SC}}(\hbar\Omega_q - \mu_s)\right), \\
I_{\text{FI}}^{(2)} &= -4\pi\hbar^2 |W|^2\nu_0^2V_{\text{SC}}^2\sum_{q} (1+2|v_q|^2)|v_q|^2\nonumber\\
&\times\int_{-\infty}^\infty \nu(E_k-h)\nu(E_k + \hbar\Omega_q + h)F(E_k,-h)\nonumber \\
&\times\left[f(E_k + \hbar\Omega_q + \mu_\downarrow) - f(E_k + \mu_\uparrow)\right]\nonumber\\
&\times\left( n_{\text{FI}}(\hbar\Omega_q - \mu_m) - n_{\text{SC}}(\hbar\Omega_q + \mu_s)\right).
\end{align}
Notice that $I_{\text{FI}}^{(1)}$ and $I_{\text{FI}}^{(2)}$ have the same sign, as they both correspond to a process in which a magnon is annihilated (or created). On the other hand, $I_{\text{SC}}^{(1)}$ and $I_{\text{SC}}^{(2)}$ have opposite signs, since the former corresponds to the addition of spin $\downarrow$, and the latter of spin $\uparrow$ (or vice versa). 
We define the spin current passing between the two materials as $I_s = \left(I_{\text{SC}} - I_{\text{FI}}\right)/2$. Doing so, we find that $I_s^{(2)} =I_{\text{SC}}^{(2)}-I_{\text{FI}}^{(2)}$ only gives a contribution on the order of $|v_q|^4$, and as $|v_q|$ is typically a small quantity, we neglect this contribution. We are thus left with $I_s = I_s^{(1)} = \left(I_{\text{SC}}^{(1)}-I_{\text{FI}}^{(1)}\right)/2$, giving
\begin{align}
I_s &= 4\pi\hbar^2|W|^2\nu_0^2V_{\text{SC}}^2\sum_q(1+|v_q|^2)u_q^2\nonumber\\
&\times\int_{-\infty}^\infty dE_k\;\nu( E_k+h)\nu(E_k + \hbar\Omega_q -h)F(E_k,+h) \nonumber \\
&\times \left[f(E_k + \hbar\Omega_q - \mu_\downarrow) - f(E_k - \mu_\uparrow)\right]\nonumber\\
&\times\left( n_{\text{FI}}(\hbar\Omega_q - \mu_m) - n_{\text{SC}}(\hbar\Omega_q - \mu_s)\right), \nonumber\\
&\equiv4\pi\hbar^2|W|^2\nu_0^2V_{\text{SC}}^2\sum_q(1+|v_q|^2)u_q^2\chi(\hbar\Omega_q,h,\mu_s)\nonumber\\
&\times\left( n_{\text{FI}}(\hbar\Omega_q - \mu_m) - n_{\text{SC}}(\hbar\Omega_q - \mu_s)\right).
\end{align}

The energy current operator on the superconductor side is given by
\begin{align}
\hat{\dot{E}}_{\text{SC}} = \frac{1}{i\hbar}\left[H_{\text{SC}},H_{\text{int}}\right],
\end{align}
whereas on the ferromagnet side it is given as
\begin{align}
\hat{\dot{E}}_{\text{FI}} = \frac{1}{i\hbar}\left[H_{\text{FI}},H_{\text{int}}\right].
\end{align}
Repeating the exact same steps as for the spin current, it is found that 
\begin{align}
\dot{E}_{\text{SC}}^{(1)} &= -\dot{E}_{\text{FI}}^{(1)} = -4\pi\hbar |W|^2\nu_0^2V_{\text{SC}}^2\sum_{q} \hbar\Omega_qu_q^2\chi(\hbar\Omega_q,h,\mu_s)\nonumber\\
&\times\left( n_{\text{FI}}(\hbar\Omega_q - \mu_m) - n_{\text{SC}}(\hbar\Omega_q - \mu_s)\right).
\end{align}
Similarly,
\begin{align}
&\dot{E}_{\text{SC}}^{(2)} = -\dot{E}_{\text{FI}}^{(2)} = -4\pi\hbar |W|^2\nu_0^2V_{\text{SC}}^2\sum_{q} \hbar\Omega_q|v_q|^2\nonumber\\
&\times\chi(\hbar\Omega_q,-h,-\mu_s)
\left( n_{\text{FI}}(\hbar\Omega_q - \mu_m) - n_{\text{SC}}(\hbar\Omega_q + \mu_s)\right).
\end{align}
Hence, $\dot{E}_{\text{SC}} + \dot{E}_{\text{FI}} = 0$, meaning that the energy current remains conserved, which is reasonable as we have introduced no inelastic scattering processes.

The heat currents are found by making the replacements $\hbar\Omega_q\to \hbar\Omega_q - \mu_m$ in $E_{\text{FI}}^{(j)}$, $\hbar\Omega_q\to \hbar\Omega_q - \mu_s$ in $E_{\text{SC}}^{(1)}$, and $\hbar\Omega_q\to \hbar\Omega_q + \mu_s$ in $E_{\text{SC}}^{(2)}$. As before, we define the average heat current as $\dot{Q} = \left(\dot{Q}_{\text{SC}} - \dot{Q}_{\text{FI}}\right)/2$, giving
\begin{align}
\dot{Q} &= -4\pi\hbar |W|^2\nu_0^2V_{\text{SC}}^2\sum_{q} \left(\hbar\Omega_q-\bar{\mu}\right)u_q^2\chi(\hbar\Omega_q,h,\mu_s)\nonumber\\
&\times\left( n_{\text{FI}}(\hbar\Omega_q - \mu_m) - n_{\text{SC}}(\hbar\Omega_q - \mu_s)\right) \nonumber \\
&-4\pi\hbar |W|^2\nu_0^2V_{\text{SC}}^2\sum_{q} \left(\hbar\Omega_q + \frac{1}{2}\Delta\mu\right)|v_q|^2\chi(\hbar\Omega_q,-h,-\mu_s)\nonumber\\
&\times\left( n_{\text{FI}}(\hbar\Omega_q - \mu_m) - n_{\text{SC}}(\hbar\Omega_q + \mu_s)\right).
\end{align}

In the linear-response regime, we get
\begin{align}
\begin{pmatrix} I_s \\ \dot{Q} \end{pmatrix} = \begin{pmatrix} G & \alpha_I \\ \alpha_Q & \kappa T \end{pmatrix} \begin{pmatrix}\Delta\mu \\ \Delta T/T \end{pmatrix},
\end{align}
with
\begin{widetext}
\begin{align}
G &= 4\pi\hbar^2|W|^2\nu_0^2V_{\text{SC}}^2\sum_q\left(1+|v_q|^2\right)u_q^2\frac{\chi(\hbar\Omega_q,h,\bar{\mu})}{4k_{\text{B}}T\sinh^2\frac{\hbar\Omega_q-\bar{\mu}}{2k_{\text{B}}T}}, \\
\alpha_I &=4\pi\hbar^2|W|^2\nu_0^2V_{\text{SC}}^2\sum_q\left(\hbar\Omega_q-\bar{\mu}\right)\left(1+|v_q|^2\right)u_q^2\frac{\chi(\hbar\Omega_q,h,\bar{\mu})}{4k_{\text{B}}T\sinh^2\frac{\hbar\Omega_q-\bar{\mu}}{2k_{\text{B}}T}}, \\
\alpha_Q &=-4\pi\hbar|W|^2\nu_0^2V_{\text{SC}}^2\sum_q\left(\hbar\Omega_q - \bar{\mu}\right)u_q^2\frac{\chi(\hbar\Omega_q,h,\bar{\mu})}{4k_{\text{B}}T\sinh^2\frac{\hbar\Omega_q-\bar{\mu}}{2k_{\text{B}}T}}, \nonumber \\
&-4\pi\hbar|W|^2\nu_0^2V_{\text{SC}}^2\sum_q\hbar\Omega_q |v_q|^2\frac{\chi(\hbar\Omega_q,-h,-\bar{\mu})\left(\sinh^2\frac{\hbar\Omega_q+\bar{\mu}}{2k_{\text{B}}T} -\sinh^2\frac{\hbar\Omega-\bar{\mu}}{2k_{\text{B}}T}\right)}{8k_{\text{B}}T\sinh^2\frac{\hbar\Omega-\bar{\mu}}{2k_{\text{B}}T}\sinh^2\frac{\hbar\Omega_q+\bar{\mu}}{2k_{\text{B}}T}}, \\
\kappa &=-4\pi\hbar|W|^2\nu_0^2V_{\text{SC}}^2\sum_q\left(\hbar\Omega_q - \bar{\mu}\right)^2u_q^2\frac{\chi(\hbar\Omega_q,h,\bar{\mu})}{4k_{\text{B}}T^2\sinh^2\frac{\hbar\Omega_q-\bar{\mu}}{2k_{\text{B}}T}}, \nonumber \\
&-4\pi\hbar|W|^2\nu_0^2V_{\text{SC}}^2\sum_q\hbar\Omega_q |v_q|^2\frac{\chi(\hbar\Omega_q,-h,-\bar{\mu})\left[\left(\hbar\Omega_q-\bar{\mu}\right)\sinh^2\frac{\hbar\Omega_q+\bar{\mu}}{2k_{\text{B}}T} +\left(\hbar\Omega_q+\bar{\mu}\right)\sinh^2\frac{\hbar\Omega_q-\bar{\mu}}{2k_{\text{B}}T}\right]}{8k_{\text{B}}T^2\sinh^2\frac{\hbar\Omega_q-\bar{\mu}}{2k_{\text{B}}T}\sinh^2\frac{\hbar\Omega_q+\bar{\mu}}{2k_{\text{B}}T}}.
\end{align}
\end{widetext}
In the limit of no dipole-dipole interaction, $u_q = 1$ and $v_q = 0$, we see that these expressions reduce to the results presented in the main text.


\bibliography{References}

\begin{thebibliography}{71}%
\makeatletter
\providecommand \@ifxundefined [1]{%
 \@ifx{#1\undefined}
}%
\providecommand \@ifnum [1]{%
 \ifnum #1\expandafter \@firstoftwo
 \else \expandafter \@secondoftwo
 \fi
}%
\providecommand \@ifx [1]{%
 \ifx #1\expandafter \@firstoftwo
 \else \expandafter \@secondoftwo
 \fi
}%
\providecommand \natexlab [1]{#1}%
\providecommand \enquote  [1]{``#1''}%
\providecommand \bibnamefont  [1]{#1}%
\providecommand \bibfnamefont [1]{#1}%
\providecommand \citenamefont [1]{#1}%
\providecommand \href@noop [0]{\@secondoftwo}%
\providecommand \href [0]{\begingroup \@sanitize@url \@href}%
\providecommand \@href[1]{\@@startlink{#1}\@@href}%
\providecommand \@@href[1]{\endgroup#1\@@endlink}%
\providecommand \@sanitize@url [0]{\catcode `\\12\catcode `\$12\catcode
  `\&12\catcode `\#12\catcode `\^12\catcode `\_12\catcode `\%12\relax}%
\providecommand \@@startlink[1]{}%
\providecommand \@@endlink[0]{}%
\providecommand \url  [0]{\begingroup\@sanitize@url \@url }%
\providecommand \@url [1]{\endgroup\@href {#1}{\urlprefix }}%
\providecommand \urlprefix  [0]{URL }%
\providecommand \Eprint [0]{\href }%
\providecommand \doibase [0]{https://doi.org/}%
\providecommand \selectlanguage [0]{\@gobble}%
\providecommand \bibinfo  [0]{\@secondoftwo}%
\providecommand \bibfield  [0]{\@secondoftwo}%
\providecommand \translation [1]{[#1]}%
\providecommand \BibitemOpen [0]{}%
\providecommand \bibitemStop [0]{}%
\providecommand \bibitemNoStop [0]{.\EOS\space}%
\providecommand \EOS [0]{\spacefactor3000\relax}%
\providecommand \BibitemShut  [1]{\csname bibitem#1\endcsname}%
\let\auto@bib@innerbib\@empty
\bibitem [{\citenamefont {Demokritov}\ \emph {et~al.}(2006)\citenamefont
  {Demokritov}, \citenamefont {Demidov}, \citenamefont {Dzyapko}, \citenamefont
  {Melkov}, \citenamefont {Serga}, \citenamefont {Hillebrands},\ and\
  \citenamefont {Slavin}}]{Demokritov2006}%
  \BibitemOpen
  \bibfield  {author} {\bibinfo {author} {\bibfnamefont {S.~O.}\ \bibnamefont
  {Demokritov}}, \bibinfo {author} {\bibfnamefont {V.~E.}\ \bibnamefont
  {Demidov}}, \bibinfo {author} {\bibfnamefont {O.}~\bibnamefont {Dzyapko}},
  \bibinfo {author} {\bibfnamefont {G.~A.}\ \bibnamefont {Melkov}}, \bibinfo
  {author} {\bibfnamefont {A.~A.}\ \bibnamefont {Serga}}, \bibinfo {author}
  {\bibfnamefont {B.}~\bibnamefont {Hillebrands}},\ and\ \bibinfo {author}
  {\bibfnamefont {A.~N.}\ \bibnamefont {Slavin}},\ }\bibfield  {title}
  {\bibinfo {title} {{Bose–Einstein condensation of quasi-equilibrium magnons
  at room temperature under pumping}},\ }\href
  {https://doi.org/10.1038/nature05117} {\bibfield  {journal} {\bibinfo
  {journal} {Nature 2006 443:7110}\ }\textbf {\bibinfo {volume} {443}},\
  \bibinfo {pages} {430} (\bibinfo {year} {2006})}\BibitemShut {NoStop}%
\bibitem [{\citenamefont {Demidov}\ \emph {et~al.}(2008)\citenamefont
  {Demidov}, \citenamefont {Dzyapko}, \citenamefont {Demokritov}, \citenamefont
  {Melkov},\ and\ \citenamefont {Slavin}}]{Demidov2008}%
  \BibitemOpen
  \bibfield  {author} {\bibinfo {author} {\bibfnamefont {V.~E.}\ \bibnamefont
  {Demidov}}, \bibinfo {author} {\bibfnamefont {O.}~\bibnamefont {Dzyapko}},
  \bibinfo {author} {\bibfnamefont {S.~O.}\ \bibnamefont {Demokritov}},
  \bibinfo {author} {\bibfnamefont {G.~A.}\ \bibnamefont {Melkov}},\ and\
  \bibinfo {author} {\bibfnamefont {A.~N.}\ \bibnamefont {Slavin}},\ }\bibfield
   {title} {\bibinfo {title} {{Observation of Spontaneous Coherence in
  Bose-Einstein Condensate of Magnons}},\ }\href
  {https://doi.org/10.1103/PhysRevLett.100.047205} {\bibfield  {journal}
  {\bibinfo  {journal} {Physical Review Letters}\ }\textbf {\bibinfo {volume}
  {100}},\ \bibinfo {pages} {047205} (\bibinfo {year} {2008})}\BibitemShut
  {NoStop}%
\bibitem [{\citenamefont {Bender}\ \emph {et~al.}(2012)\citenamefont {Bender},
  \citenamefont {Duine},\ and\ \citenamefont {Tserkovnyak}}]{Bender2012}%
  \BibitemOpen
  \bibfield  {author} {\bibinfo {author} {\bibfnamefont {S.~A.}\ \bibnamefont
  {Bender}}, \bibinfo {author} {\bibfnamefont {R.~A.}\ \bibnamefont {Duine}},\
  and\ \bibinfo {author} {\bibfnamefont {Y.}~\bibnamefont {Tserkovnyak}},\
  }\bibfield  {title} {\bibinfo {title} {Electronic pumping of quasiequilibrium
  bose-einstein-condensed magnons},\ }\href
  {https://doi.org/10.1103/PhysRevLett.108.246601} {\bibfield  {journal}
  {\bibinfo  {journal} {Physical Review Letters}\ }\textbf {\bibinfo {volume}
  {108}},\ \bibinfo {pages} {246601} (\bibinfo {year} {2012})}\BibitemShut
  {NoStop}%
\bibitem [{\citenamefont {Bauer}\ \emph {et~al.}(2012)\citenamefont {Bauer},
  \citenamefont {Saitoh},\ and\ \citenamefont {van Wees}}]{Bauer2012}%
  \BibitemOpen
  \bibfield  {author} {\bibinfo {author} {\bibfnamefont {G.~E.~W.}\
  \bibnamefont {Bauer}}, \bibinfo {author} {\bibfnamefont {E.}~\bibnamefont
  {Saitoh}},\ and\ \bibinfo {author} {\bibfnamefont {B.~J.}\ \bibnamefont {van
  Wees}},\ }\bibfield  {title} {\bibinfo {title} {{Spin caloritronics}},\
  }\href@noop {} {\bibfield  {journal} {\bibinfo  {journal} {Nature Materials}\
  }\textbf {\bibinfo {volume} {11}},\ \bibinfo {pages} {391} (\bibinfo {year}
  {2012})}\BibitemShut {NoStop}%
\bibitem [{\citenamefont {Kruglyak}\ \emph {et~al.}(2010)\citenamefont
  {Kruglyak}, \citenamefont {Demokritov},\ and\ \citenamefont
  {Grundler}}]{Kruglyak2010}%
  \BibitemOpen
  \bibfield  {author} {\bibinfo {author} {\bibfnamefont {V.~V.}\ \bibnamefont
  {Kruglyak}}, \bibinfo {author} {\bibfnamefont {S.~O.}\ \bibnamefont
  {Demokritov}},\ and\ \bibinfo {author} {\bibfnamefont {D.}~\bibnamefont
  {Grundler}},\ }\bibfield  {title} {\bibinfo {title} {{Magnonics}},\ }\href
  {https://doi.org/10.1088/0022-3727/43/26/264001} {\bibfield  {journal}
  {\bibinfo  {journal} {Journal of Physics D: Applied Physics}\ }\textbf
  {\bibinfo {volume} {43}},\ \bibinfo {pages} {264001} (\bibinfo {year}
  {2010})}\BibitemShut {NoStop}%
\bibitem [{\citenamefont {Serga}\ \emph {et~al.}(2010)\citenamefont {Serga},
  \citenamefont {Chumak},\ and\ \citenamefont {Hillebrands}}]{Serga2010}%
  \BibitemOpen
  \bibfield  {author} {\bibinfo {author} {\bibfnamefont {A.~A.}\ \bibnamefont
  {Serga}}, \bibinfo {author} {\bibfnamefont {A.~V.}\ \bibnamefont {Chumak}},\
  and\ \bibinfo {author} {\bibfnamefont {B.}~\bibnamefont {Hillebrands}},\
  }\bibfield  {title} {\bibinfo {title} {{YIG magnonics}},\ }\href
  {https://doi.org/10.1088/0022-3727/43/26/264002} {\bibfield  {journal}
  {\bibinfo  {journal} {Journal of Physics D: Applied Physics}\ }\textbf
  {\bibinfo {volume} {43}},\ \bibinfo {pages} {264002} (\bibinfo {year}
  {2010})}\BibitemShut {NoStop}%
\bibitem [{\citenamefont {Chumak}\ \emph {et~al.}(2015)\citenamefont {Chumak},
  \citenamefont {Vasyuchka}, \citenamefont {Serga},\ and\ \citenamefont
  {Hillebrands}}]{Chumak2015}%
  \BibitemOpen
  \bibfield  {author} {\bibinfo {author} {\bibfnamefont {A.~V.}\ \bibnamefont
  {Chumak}}, \bibinfo {author} {\bibfnamefont {V.}~\bibnamefont {Vasyuchka}},
  \bibinfo {author} {\bibfnamefont {A.}~\bibnamefont {Serga}},\ and\ \bibinfo
  {author} {\bibfnamefont {B.}~\bibnamefont {Hillebrands}},\ }\bibfield
  {title} {\bibinfo {title} {{Magnon spintronics}},\ }\href
  {https://doi.org/10.1038/nphys3347} {\bibfield  {journal} {\bibinfo
  {journal} {Nature Physics 2014 11:6}\ }\textbf {\bibinfo {volume} {11}},\
  \bibinfo {pages} {453} (\bibinfo {year} {2015})}\BibitemShut {NoStop}%
\bibitem [{\citenamefont {Chumak}\ and\ \citenamefont
  {Schultheiss}(2017)}]{Chumak2017}%
  \BibitemOpen
  \bibfield  {author} {\bibinfo {author} {\bibfnamefont {A.~V.}\ \bibnamefont
  {Chumak}}\ and\ \bibinfo {author} {\bibfnamefont {H.}~\bibnamefont
  {Schultheiss}},\ }\bibfield  {title} {\bibinfo {title} {{Magnonics: spin
  waves connecting charges, spins and photons}},\ }\href
  {https://doi.org/10.1088/1361-6463/AA7715} {\bibfield  {journal} {\bibinfo
  {journal} {Journal of Physics D: Applied Physics}\ }\textbf {\bibinfo
  {volume} {50}},\ \bibinfo {pages} {300201} (\bibinfo {year}
  {2017})}\BibitemShut {NoStop}%
\bibitem [{\citenamefont {Althammer}(2018)}]{Althammer2018}%
  \BibitemOpen
  \bibfield  {author} {\bibinfo {author} {\bibfnamefont {M.}~\bibnamefont
  {Althammer}},\ }\bibfield  {title} {\bibinfo {title} {{Pure spin currents in
  magnetically ordered insulator/normal metal heterostructures}},\ }\href
  {https://doi.org/10.1088/1361-6463/AACA89} {\bibfield  {journal} {\bibinfo
  {journal} {Journal of Physics D: Applied Physics}\ }\textbf {\bibinfo
  {volume} {51}},\ \bibinfo {pages} {313001} (\bibinfo {year}
  {2018})}\BibitemShut {NoStop}%
\bibitem [{\citenamefont {Althammer}(2021)}]{Althammer2021}%
  \BibitemOpen
  \bibfield  {author} {\bibinfo {author} {\bibfnamefont {M.}~\bibnamefont
  {Althammer}},\ }\bibfield  {title} {\bibinfo {title} {{All-Electrical Magnon
  Transport Experiments in Magnetically Ordered Insulators}},\ }\href
  {https://doi.org/10.1002/PSSR.202100130} {\bibfield  {journal} {\bibinfo
  {journal} {physica status solidi (RRL) – Rapid Research Letters}\ }\textbf
  {\bibinfo {volume} {15}},\ \bibinfo {pages} {2100130} (\bibinfo {year}
  {2021})}\BibitemShut {NoStop}%
\bibitem [{\citenamefont {Pirro}\ \emph {et~al.}(2021)\citenamefont {Pirro},
  \citenamefont {Vasyuchka}, \citenamefont {Serga},\ and\ \citenamefont
  {Hillebrands}}]{Pirro2021}%
  \BibitemOpen
  \bibfield  {author} {\bibinfo {author} {\bibfnamefont {P.}~\bibnamefont
  {Pirro}}, \bibinfo {author} {\bibfnamefont {V.~I.}\ \bibnamefont
  {Vasyuchka}}, \bibinfo {author} {\bibfnamefont {A.~A.}\ \bibnamefont
  {Serga}},\ and\ \bibinfo {author} {\bibfnamefont {B.}~\bibnamefont
  {Hillebrands}},\ }\bibfield  {title} {\bibinfo {title} {{Advances in coherent
  magnonics}},\ }\href {https://doi.org/10.1038/s41578-021-00332-w} {\bibfield
  {journal} {\bibinfo  {journal} {Nature Reviews Materials 2021}\ ,\ \bibinfo
  {pages} {1}} (\bibinfo {year} {2021})}\BibitemShut {NoStop}%
\bibitem [{\citenamefont {Barman}\ \emph {et~al.}(2021)\citenamefont {Barman},
  \citenamefont {Gubbiotti}, \citenamefont {Ladak}, \citenamefont {Adeyeye},
  \citenamefont {Krawczyk}, \citenamefont {Gr{\"{a}}fe}, \citenamefont
  {Adelmann}, \citenamefont {Cotofana}, \citenamefont {Naeemi}, \citenamefont
  {Vasyuchka}, \citenamefont {Hillebrands}, \citenamefont {Nikitov},
  \citenamefont {Yu}, \citenamefont {Grundler}, \citenamefont {Sadovnikov},
  \citenamefont {Grachev}, \citenamefont {Sheshukova}, \citenamefont
  {Duquesne}, \citenamefont {Marangolo}, \citenamefont {Csaba}, \citenamefont
  {Porod}, \citenamefont {Demidov}, \citenamefont {Urazhdin}, \citenamefont
  {Demokritov}, \citenamefont {Albisetti}, \citenamefont {Petti}, \citenamefont
  {Bertacco}, \citenamefont {Schultheiss}, \citenamefont {Kruglyak},
  \citenamefont {Poimanov}, \citenamefont {Sahoo}, \citenamefont {Sinha},
  \citenamefont {Yang}, \citenamefont {M{\"{u}}nzenberg}, \citenamefont
  {Moriyama}, \citenamefont {Mizukami}, \citenamefont {Landeros}, \citenamefont
  {Gallardo}, \citenamefont {Carlotti}, \citenamefont {Kim}, \citenamefont
  {Stamps}, \citenamefont {Camley}, \citenamefont {Rana}, \citenamefont
  {Otani}, \citenamefont {Yu}, \citenamefont {Yu}, \citenamefont {Bauer},
  \citenamefont {Back}, \citenamefont {Uhrig}, \citenamefont {Dobrovolskiy},
  \citenamefont {Budinska}, \citenamefont {Qin}, \citenamefont {van Dijken},
  \citenamefont {Chumak}, \citenamefont {Khitun}, \citenamefont {Nikonov},
  \citenamefont {Young}, \citenamefont {Zingsem},\ and\ \citenamefont
  {Winklhofer}}]{Barman2021}%
  \BibitemOpen
  \bibfield  {author} {\bibinfo {author} {\bibfnamefont {A.}~\bibnamefont
  {Barman}}, \bibinfo {author} {\bibfnamefont {G.}~\bibnamefont {Gubbiotti}},
  \bibinfo {author} {\bibfnamefont {S.}~\bibnamefont {Ladak}}, \bibinfo
  {author} {\bibfnamefont {A.~O.}\ \bibnamefont {Adeyeye}}, \bibinfo {author}
  {\bibfnamefont {M.}~\bibnamefont {Krawczyk}}, \bibinfo {author}
  {\bibfnamefont {J.}~\bibnamefont {Gr{\"{a}}fe}}, \bibinfo {author}
  {\bibfnamefont {C.}~\bibnamefont {Adelmann}}, \bibinfo {author}
  {\bibfnamefont {S.}~\bibnamefont {Cotofana}}, \bibinfo {author}
  {\bibfnamefont {A.}~\bibnamefont {Naeemi}}, \bibinfo {author} {\bibfnamefont
  {V.~I.}\ \bibnamefont {Vasyuchka}}, \bibinfo {author} {\bibfnamefont
  {B.}~\bibnamefont {Hillebrands}}, \bibinfo {author} {\bibfnamefont {S.~A.}\
  \bibnamefont {Nikitov}}, \bibinfo {author} {\bibfnamefont {H.}~\bibnamefont
  {Yu}}, \bibinfo {author} {\bibfnamefont {D.}~\bibnamefont {Grundler}},
  \bibinfo {author} {\bibfnamefont {A.~V.}\ \bibnamefont {Sadovnikov}},
  \bibinfo {author} {\bibfnamefont {A.~A.}\ \bibnamefont {Grachev}}, \bibinfo
  {author} {\bibfnamefont {S.~E.}\ \bibnamefont {Sheshukova}}, \bibinfo
  {author} {\bibfnamefont {J.-Y.}\ \bibnamefont {Duquesne}}, \bibinfo {author}
  {\bibfnamefont {M.}~\bibnamefont {Marangolo}}, \bibinfo {author}
  {\bibfnamefont {G.}~\bibnamefont {Csaba}}, \bibinfo {author} {\bibfnamefont
  {W.}~\bibnamefont {Porod}}, \bibinfo {author} {\bibfnamefont {V.~E.}\
  \bibnamefont {Demidov}}, \bibinfo {author} {\bibfnamefont {S.}~\bibnamefont
  {Urazhdin}}, \bibinfo {author} {\bibfnamefont {S.~O.}\ \bibnamefont
  {Demokritov}}, \bibinfo {author} {\bibfnamefont {E.}~\bibnamefont
  {Albisetti}}, \bibinfo {author} {\bibfnamefont {D.}~\bibnamefont {Petti}},
  \bibinfo {author} {\bibfnamefont {R.}~\bibnamefont {Bertacco}}, \bibinfo
  {author} {\bibfnamefont {H.}~\bibnamefont {Schultheiss}}, \bibinfo {author}
  {\bibfnamefont {V.~V.}\ \bibnamefont {Kruglyak}}, \bibinfo {author}
  {\bibfnamefont {V.~D.}\ \bibnamefont {Poimanov}}, \bibinfo {author}
  {\bibfnamefont {S.}~\bibnamefont {Sahoo}}, \bibinfo {author} {\bibfnamefont
  {J.}~\bibnamefont {Sinha}}, \bibinfo {author} {\bibfnamefont
  {H.}~\bibnamefont {Yang}}, \bibinfo {author} {\bibfnamefont {M.}~\bibnamefont
  {M{\"{u}}nzenberg}}, \bibinfo {author} {\bibfnamefont {T.}~\bibnamefont
  {Moriyama}}, \bibinfo {author} {\bibfnamefont {S.}~\bibnamefont {Mizukami}},
  \bibinfo {author} {\bibfnamefont {P.}~\bibnamefont {Landeros}}, \bibinfo
  {author} {\bibfnamefont {R.~A.}\ \bibnamefont {Gallardo}}, \bibinfo {author}
  {\bibfnamefont {G.}~\bibnamefont {Carlotti}}, \bibinfo {author}
  {\bibfnamefont {J.-V.}\ \bibnamefont {Kim}}, \bibinfo {author} {\bibfnamefont
  {R.~L.}\ \bibnamefont {Stamps}}, \bibinfo {author} {\bibfnamefont {R.~E.}\
  \bibnamefont {Camley}}, \bibinfo {author} {\bibfnamefont {B.}~\bibnamefont
  {Rana}}, \bibinfo {author} {\bibfnamefont {Y.}~\bibnamefont {Otani}},
  \bibinfo {author} {\bibfnamefont {W.}~\bibnamefont {Yu}}, \bibinfo {author}
  {\bibfnamefont {T.}~\bibnamefont {Yu}}, \bibinfo {author} {\bibfnamefont
  {G.~E.~W.}\ \bibnamefont {Bauer}}, \bibinfo {author} {\bibfnamefont
  {C.}~\bibnamefont {Back}}, \bibinfo {author} {\bibfnamefont {G.~S.}\
  \bibnamefont {Uhrig}}, \bibinfo {author} {\bibfnamefont {O.~V.}\ \bibnamefont
  {Dobrovolskiy}}, \bibinfo {author} {\bibfnamefont {B.}~\bibnamefont
  {Budinska}}, \bibinfo {author} {\bibfnamefont {H.}~\bibnamefont {Qin}},
  \bibinfo {author} {\bibfnamefont {S.}~\bibnamefont {van Dijken}}, \bibinfo
  {author} {\bibfnamefont {A.~V.}\ \bibnamefont {Chumak}}, \bibinfo {author}
  {\bibfnamefont {A.}~\bibnamefont {Khitun}}, \bibinfo {author} {\bibfnamefont
  {D.~E.}\ \bibnamefont {Nikonov}}, \bibinfo {author} {\bibfnamefont {I.~A.}\
  \bibnamefont {Young}}, \bibinfo {author} {\bibfnamefont {B.~W.}\ \bibnamefont
  {Zingsem}},\ and\ \bibinfo {author} {\bibfnamefont {M.}~\bibnamefont
  {Winklhofer}},\ }\bibfield  {title} {\bibinfo {title} {{The 2021 Magnonics
  Roadmap}},\ }\href {https://doi.org/10.1088/1361-648X/ABEC1A} {\bibfield
  {journal} {\bibinfo  {journal} {Journal of Physics: Condensed Matter}\
  }\textbf {\bibinfo {volume} {33}},\ \bibinfo {pages} {413001} (\bibinfo
  {year} {2021})}\BibitemShut {NoStop}%
\bibitem [{\citenamefont {Nakata}\ \emph {et~al.}(2017)\citenamefont {Nakata},
  \citenamefont {Simon},\ and\ \citenamefont {Loss}}]{Nakata2017}%
  \BibitemOpen
  \bibfield  {author} {\bibinfo {author} {\bibfnamefont {K.}~\bibnamefont
  {Nakata}}, \bibinfo {author} {\bibfnamefont {P.}~\bibnamefont {Simon}},\ and\
  \bibinfo {author} {\bibfnamefont {D.}~\bibnamefont {Loss}},\ }\bibfield
  {title} {\bibinfo {title} {Spin currents and magnon dynamics in insulating
  magnets},\ }\href {https://doi.org/10.1088/1361-6463/aa5b09} {\bibfield
  {journal} {\bibinfo  {journal} {Journal of Physics D: Applied Physics}\
  }\textbf {\bibinfo {volume} {50}},\ \bibinfo {pages} {114004} (\bibinfo
  {year} {2017})}\BibitemShut {NoStop}%
\bibitem [{\citenamefont {Cornelissen}\ \emph {et~al.}(2016)\citenamefont
  {Cornelissen}, \citenamefont {Peters}, \citenamefont {Bauer}, \citenamefont
  {Duine},\ and\ \citenamefont {van Wees}}]{Cornelissen2016}%
  \BibitemOpen
  \bibfield  {author} {\bibinfo {author} {\bibfnamefont {L.~J.}\ \bibnamefont
  {Cornelissen}}, \bibinfo {author} {\bibfnamefont {K.~J.~H.}\ \bibnamefont
  {Peters}}, \bibinfo {author} {\bibfnamefont {G.~E.~W.}\ \bibnamefont
  {Bauer}}, \bibinfo {author} {\bibfnamefont {R.~A.}\ \bibnamefont {Duine}},\
  and\ \bibinfo {author} {\bibfnamefont {B.~J.}\ \bibnamefont {van Wees}},\
  }\bibfield  {title} {\bibinfo {title} {{Magnon spin transport driven by the
  magnon chemical potential in a magnetic insulator}},\ }\href
  {https://doi.org/10.1103/PhysRevB.94.014412} {\bibfield  {journal} {\bibinfo
  {journal} {Physical Review B}\ }\textbf {\bibinfo {volume} {94}},\ \bibinfo
  {pages} {014412} (\bibinfo {year} {2016})}\BibitemShut {NoStop}%
\bibitem [{\citenamefont {Kamra}\ and\ \citenamefont
  {Belzig}(2016)}]{Kamra2016}%
  \BibitemOpen
  \bibfield  {author} {\bibinfo {author} {\bibfnamefont {A.}~\bibnamefont
  {Kamra}}\ and\ \bibinfo {author} {\bibfnamefont {W.}~\bibnamefont {Belzig}},\
  }\bibfield  {title} {\bibinfo {title} {{Magnon-mediated spin current noise in
  ferromagnet | nonmagnetic conductor hybrids}},\ }\href
  {https://doi.org/10.1103/PhysRevB.94.014419} {\bibfield  {journal} {\bibinfo
  {journal} {Physical Review B}\ }\textbf {\bibinfo {volume} {94}},\ \bibinfo
  {pages} {014419} (\bibinfo {year} {2016})}\BibitemShut {NoStop}%
\bibitem [{\citenamefont {Olsson}\ \emph {et~al.}(2020)\citenamefont {Olsson},
  \citenamefont {An}, \citenamefont {Fiete}, \citenamefont {Zhou},
  \citenamefont {Shi},\ and\ \citenamefont {Li}}]{Olsson2020}%
  \BibitemOpen
  \bibfield  {author} {\bibinfo {author} {\bibfnamefont {K.~S.}\ \bibnamefont
  {Olsson}}, \bibinfo {author} {\bibfnamefont {K.}~\bibnamefont {An}}, \bibinfo
  {author} {\bibfnamefont {G.~A.}\ \bibnamefont {Fiete}}, \bibinfo {author}
  {\bibfnamefont {J.}~\bibnamefont {Zhou}}, \bibinfo {author} {\bibfnamefont
  {L.}~\bibnamefont {Shi}},\ and\ \bibinfo {author} {\bibfnamefont
  {X.}~\bibnamefont {Li}},\ }\bibfield  {title} {\bibinfo {title} {{Pure Spin
  Current and Magnon Chemical Potential in a Nonequilibrium Magnetic
  Insulator}},\ }\href {https://doi.org/10.1103/PhysRevX.10.021029} {\bibfield
  {journal} {\bibinfo  {journal} {Physical Review X}\ }\textbf {\bibinfo
  {volume} {10}},\ \bibinfo {pages} {021029} (\bibinfo {year}
  {2020})}\BibitemShut {NoStop}%
\bibitem [{\citenamefont {Schlitz}\ \emph {et~al.}(2021)\citenamefont
  {Schlitz}, \citenamefont {V{\'{e}}lez}, \citenamefont {Kamra}, \citenamefont
  {Lambert}, \citenamefont {Lammel}, \citenamefont {Goennenwein},\ and\
  \citenamefont {Gambardella}}]{Schlitz2021}%
  \BibitemOpen
  \bibfield  {author} {\bibinfo {author} {\bibfnamefont {R.}~\bibnamefont
  {Schlitz}}, \bibinfo {author} {\bibfnamefont {S.}~\bibnamefont
  {V{\'{e}}lez}}, \bibinfo {author} {\bibfnamefont {A.}~\bibnamefont {Kamra}},
  \bibinfo {author} {\bibfnamefont {C.-H.}\ \bibnamefont {Lambert}}, \bibinfo
  {author} {\bibfnamefont {M.}~\bibnamefont {Lammel}}, \bibinfo {author}
  {\bibfnamefont {S.~T.}\ \bibnamefont {Goennenwein}},\ and\ \bibinfo {author}
  {\bibfnamefont {P.}~\bibnamefont {Gambardella}},\ }\bibfield  {title}
  {\bibinfo {title} {{Control of Nonlocal Magnon Spin Transport via Magnon
  Drift Currents}},\ }\href {https://doi.org/10.1103/PhysRevLett.126.257201}
  {\bibfield  {journal} {\bibinfo  {journal} {Physical Review Letters}\
  }\textbf {\bibinfo {volume} {126}},\ \bibinfo {pages} {257201} (\bibinfo
  {year} {2021})}\BibitemShut {NoStop}%
\bibitem [{\citenamefont {Du}\ \emph {et~al.}(2017)\citenamefont {Du},
  \citenamefont {van~der Sar}, \citenamefont {Zhou}, \citenamefont {Upadhyaya},
  \citenamefont {Casola}, \citenamefont {Zhang}, \citenamefont {Onbasli},
  \citenamefont {Ross}, \citenamefont {Walsworth}, \citenamefont
  {Tserkovnyak},\ and\ \citenamefont {Yacoby}}]{Du2017}%
  \BibitemOpen
  \bibfield  {author} {\bibinfo {author} {\bibfnamefont {C.}~\bibnamefont
  {Du}}, \bibinfo {author} {\bibfnamefont {T.}~\bibnamefont {van~der Sar}},
  \bibinfo {author} {\bibfnamefont {T.~X.}\ \bibnamefont {Zhou}}, \bibinfo
  {author} {\bibfnamefont {P.}~\bibnamefont {Upadhyaya}}, \bibinfo {author}
  {\bibfnamefont {F.}~\bibnamefont {Casola}}, \bibinfo {author} {\bibfnamefont
  {H.}~\bibnamefont {Zhang}}, \bibinfo {author} {\bibfnamefont {M.~C.}\
  \bibnamefont {Onbasli}}, \bibinfo {author} {\bibfnamefont {C.~A.}\
  \bibnamefont {Ross}}, \bibinfo {author} {\bibfnamefont {R.~L.}\ \bibnamefont
  {Walsworth}}, \bibinfo {author} {\bibfnamefont {Y.}~\bibnamefont
  {Tserkovnyak}},\ and\ \bibinfo {author} {\bibfnamefont {A.}~\bibnamefont
  {Yacoby}},\ }\bibfield  {title} {\bibinfo {title} {Control and local
  measurement of the spin chemical potential in a magnetic insulator},\ }\href
  {https://doi.org/10.1126/science.aak9611} {\bibfield  {journal} {\bibinfo
  {journal} {Science}\ }\textbf {\bibinfo {volume} {357}},\ \bibinfo {pages}
  {195} (\bibinfo {year} {2017})}\BibitemShut {NoStop}%
\bibitem [{\citenamefont {Kajiwara}\ \emph {et~al.}(2010)\citenamefont
  {Kajiwara}, \citenamefont {Harii}, \citenamefont {Takahashi}, \citenamefont
  {Ohe}, \citenamefont {Uchida}, \citenamefont {Mizuguchi}, \citenamefont
  {Umezawa}, \citenamefont {Kawai}, \citenamefont {Ando}, \citenamefont
  {Takanashi}, \citenamefont {Maekawa},\ and\ \citenamefont
  {Saitoh}}]{Kajiwara2010a}%
  \BibitemOpen
  \bibfield  {author} {\bibinfo {author} {\bibfnamefont {Y.}~\bibnamefont
  {Kajiwara}}, \bibinfo {author} {\bibfnamefont {K.}~\bibnamefont {Harii}},
  \bibinfo {author} {\bibfnamefont {S.}~\bibnamefont {Takahashi}}, \bibinfo
  {author} {\bibfnamefont {J.}~\bibnamefont {Ohe}}, \bibinfo {author}
  {\bibfnamefont {K.}~\bibnamefont {Uchida}}, \bibinfo {author} {\bibfnamefont
  {M.}~\bibnamefont {Mizuguchi}}, \bibinfo {author} {\bibfnamefont
  {H.}~\bibnamefont {Umezawa}}, \bibinfo {author} {\bibfnamefont
  {H.}~\bibnamefont {Kawai}}, \bibinfo {author} {\bibfnamefont
  {K.}~\bibnamefont {Ando}}, \bibinfo {author} {\bibfnamefont {K.}~\bibnamefont
  {Takanashi}}, \bibinfo {author} {\bibfnamefont {S.}~\bibnamefont {Maekawa}},\
  and\ \bibinfo {author} {\bibfnamefont {E.}~\bibnamefont {Saitoh}},\
  }\bibfield  {title} {\bibinfo {title} {{Transmission of electrical signals by
  spin-wave interconversion in a magnetic insulator}},\ }\href
  {https://doi.org/10.1038/nature08876} {\bibfield  {journal} {\bibinfo
  {journal} {Nature 2010 464:7286}\ }\textbf {\bibinfo {volume} {464}},\
  \bibinfo {pages} {262} (\bibinfo {year} {2010})}\BibitemShut {NoStop}%
\bibitem [{\citenamefont {Cornelissen}\ \emph {et~al.}(2015)\citenamefont
  {Cornelissen}, \citenamefont {Liu}, \citenamefont {Duine}, \citenamefont
  {Youssef},\ and\ \citenamefont {van Wees}}]{Cornelissen2015}%
  \BibitemOpen
  \bibfield  {author} {\bibinfo {author} {\bibfnamefont {L.~J.}\ \bibnamefont
  {Cornelissen}}, \bibinfo {author} {\bibfnamefont {J.}~\bibnamefont {Liu}},
  \bibinfo {author} {\bibfnamefont {R.~A.}\ \bibnamefont {Duine}}, \bibinfo
  {author} {\bibfnamefont {J.~B.}\ \bibnamefont {Youssef}},\ and\ \bibinfo
  {author} {\bibfnamefont {B.~J.}\ \bibnamefont {van Wees}},\ }\bibfield
  {title} {\bibinfo {title} {{Long-distance transport of magnon spin
  information in a magnetic insulator at room temperature}},\ }\href
  {https://doi.org/10.1038/nphys3465} {\bibfield  {journal} {\bibinfo
  {journal} {Nature Physics 2015 11:12}\ }\textbf {\bibinfo {volume} {11}},\
  \bibinfo {pages} {1022} (\bibinfo {year} {2015})}\BibitemShut {NoStop}%
\bibitem [{\citenamefont {Goennenwein}\ \emph {et~al.}(2015)\citenamefont
  {Goennenwein}, \citenamefont {Schlitz}, \citenamefont {Pernpeintner},
  \citenamefont {Ganzhorn}, \citenamefont {Althammer}, \citenamefont {Gross},\
  and\ \citenamefont {Huebl}}]{Goennenwein2015}%
  \BibitemOpen
  \bibfield  {author} {\bibinfo {author} {\bibfnamefont {S.~T.~B.}\
  \bibnamefont {Goennenwein}}, \bibinfo {author} {\bibfnamefont
  {R.}~\bibnamefont {Schlitz}}, \bibinfo {author} {\bibfnamefont
  {M.}~\bibnamefont {Pernpeintner}}, \bibinfo {author} {\bibfnamefont
  {K.}~\bibnamefont {Ganzhorn}}, \bibinfo {author} {\bibfnamefont
  {M.}~\bibnamefont {Althammer}}, \bibinfo {author} {\bibfnamefont
  {R.}~\bibnamefont {Gross}},\ and\ \bibinfo {author} {\bibfnamefont
  {H.}~\bibnamefont {Huebl}},\ }\bibfield  {title} {\bibinfo {title}
  {{Non-local magnetoresistance in YIG/Pt nanostructures}},\ }\href
  {https://doi.org/10.1063/1.4935074} {\bibfield  {journal} {\bibinfo
  {journal} {Applied Physics Letters}\ }\textbf {\bibinfo {volume} {107}},\
  \bibinfo {pages} {172405} (\bibinfo {year} {2015})}\BibitemShut {NoStop}%
\bibitem [{\citenamefont {Shan}\ \emph {et~al.}(2017)\citenamefont {Shan},
  \citenamefont {Bougiatioti}, \citenamefont {Liang}, \citenamefont {Reiss},
  \citenamefont {Kuschel},\ and\ \citenamefont {van Wees}}]{Shan2017}%
  \BibitemOpen
  \bibfield  {author} {\bibinfo {author} {\bibfnamefont {J.}~\bibnamefont
  {Shan}}, \bibinfo {author} {\bibfnamefont {P.}~\bibnamefont {Bougiatioti}},
  \bibinfo {author} {\bibfnamefont {L.}~\bibnamefont {Liang}}, \bibinfo
  {author} {\bibfnamefont {G.}~\bibnamefont {Reiss}}, \bibinfo {author}
  {\bibfnamefont {T.}~\bibnamefont {Kuschel}},\ and\ \bibinfo {author}
  {\bibfnamefont {B.~J.}\ \bibnamefont {van Wees}},\ }\bibfield  {title}
  {\bibinfo {title} {{Nonlocal magnon spin transport in NiFe2O4 thin films}},\
  }\href {https://doi.org/10.1063/1.4979408} {\bibfield  {journal} {\bibinfo
  {journal} {Applied Physics Letters}\ }\textbf {\bibinfo {volume} {110}},\
  \bibinfo {pages} {132406} (\bibinfo {year} {2017})}\BibitemShut {NoStop}%
\bibitem [{\citenamefont {Cornelissen}\ \emph {et~al.}(2018)\citenamefont
  {Cornelissen}, \citenamefont {Liu}, \citenamefont {van Wees},\ and\
  \citenamefont {Duine}}]{Cornelissen2018}%
  \BibitemOpen
  \bibfield  {author} {\bibinfo {author} {\bibfnamefont {L.}~\bibnamefont
  {Cornelissen}}, \bibinfo {author} {\bibfnamefont {J.}~\bibnamefont {Liu}},
  \bibinfo {author} {\bibfnamefont {B.}~\bibnamefont {van Wees}},\ and\
  \bibinfo {author} {\bibfnamefont {R.}~\bibnamefont {Duine}},\ }\bibfield
  {title} {\bibinfo {title} {{Spin-Current-Controlled Modulation of the Magnon
  Spin Conductance in a Three-Terminal Magnon Transistor}},\ }\href
  {https://doi.org/10.1103/PhysRevLett.120.097702} {\bibfield  {journal}
  {\bibinfo  {journal} {Physical Review Letters}\ }\textbf {\bibinfo {volume}
  {120}},\ \bibinfo {pages} {097702} (\bibinfo {year} {2018})}\BibitemShut
  {NoStop}%
\bibitem [{\citenamefont {Wimmer}\ \emph {et~al.}(2019)\citenamefont {Wimmer},
  \citenamefont {Althammer}, \citenamefont {Liensberger}, \citenamefont
  {Vlietstra}, \citenamefont {Gepr\"ags}, \citenamefont {Weiler}, \citenamefont
  {Gross},\ and\ \citenamefont {Huebl}}]{Wimmer2019}%
  \BibitemOpen
  \bibfield  {author} {\bibinfo {author} {\bibfnamefont {T.}~\bibnamefont
  {Wimmer}}, \bibinfo {author} {\bibfnamefont {M.}~\bibnamefont {Althammer}},
  \bibinfo {author} {\bibfnamefont {L.}~\bibnamefont {Liensberger}}, \bibinfo
  {author} {\bibfnamefont {N.}~\bibnamefont {Vlietstra}}, \bibinfo {author}
  {\bibfnamefont {S.}~\bibnamefont {Gepr\"ags}}, \bibinfo {author}
  {\bibfnamefont {M.}~\bibnamefont {Weiler}}, \bibinfo {author} {\bibfnamefont
  {R.}~\bibnamefont {Gross}},\ and\ \bibinfo {author} {\bibfnamefont
  {H.}~\bibnamefont {Huebl}},\ }\bibfield  {title} {\bibinfo {title} {Spin
  transport in a magnetic insulator with zero effective damping},\ }\href
  {https://doi.org/10.1103/PhysRevLett.123.257201} {\bibfield  {journal}
  {\bibinfo  {journal} {Phys. Rev. Lett.}\ }\textbf {\bibinfo {volume} {123}},\
  \bibinfo {pages} {257201} (\bibinfo {year} {2019})}\BibitemShut {NoStop}%
\bibitem [{\citenamefont {Wimmer}\ \emph {et~al.}(2020)\citenamefont {Wimmer},
  \citenamefont {Kamra}, \citenamefont {G\"uckelhorn}, \citenamefont {Opel},
  \citenamefont {Gepr\"ags}, \citenamefont {Gross}, \citenamefont {Huebl},\
  and\ \citenamefont {Althammer}}]{Wimmer2020}%
  \BibitemOpen
  \bibfield  {author} {\bibinfo {author} {\bibfnamefont {T.}~\bibnamefont
  {Wimmer}}, \bibinfo {author} {\bibfnamefont {A.}~\bibnamefont {Kamra}},
  \bibinfo {author} {\bibfnamefont {J.}~\bibnamefont {G\"uckelhorn}}, \bibinfo
  {author} {\bibfnamefont {M.}~\bibnamefont {Opel}}, \bibinfo {author}
  {\bibfnamefont {S.}~\bibnamefont {Gepr\"ags}}, \bibinfo {author}
  {\bibfnamefont {R.}~\bibnamefont {Gross}}, \bibinfo {author} {\bibfnamefont
  {H.}~\bibnamefont {Huebl}},\ and\ \bibinfo {author} {\bibfnamefont
  {M.}~\bibnamefont {Althammer}},\ }\bibfield  {title} {\bibinfo {title}
  {Observation of antiferromagnetic magnon pseudospin dynamics and the hanle
  effect},\ }\href {https://doi.org/10.1103/PhysRevLett.125.247204} {\bibfield
  {journal} {\bibinfo  {journal} {Phys. Rev. Lett.}\ }\textbf {\bibinfo
  {volume} {125}},\ \bibinfo {pages} {247204} (\bibinfo {year}
  {2020})}\BibitemShut {NoStop}%
\bibitem [{\citenamefont {An}\ \emph {et~al.}(2021)\citenamefont {An},
  \citenamefont {Kohno}, \citenamefont {Thiery}, \citenamefont {Reitz},
  \citenamefont {Vila}, \citenamefont {Naletov}, \citenamefont {Beaulieu},
  \citenamefont {Youssef}, \citenamefont {de~Loubens}, \citenamefont
  {Tserkovnyak},\ and\ \citenamefont {Klein}}]{An2021}%
  \BibitemOpen
  \bibfield  {author} {\bibinfo {author} {\bibfnamefont {K.}~\bibnamefont
  {An}}, \bibinfo {author} {\bibfnamefont {R.}~\bibnamefont {Kohno}}, \bibinfo
  {author} {\bibfnamefont {N.}~\bibnamefont {Thiery}}, \bibinfo {author}
  {\bibfnamefont {D.}~\bibnamefont {Reitz}}, \bibinfo {author} {\bibfnamefont
  {L.}~\bibnamefont {Vila}}, \bibinfo {author} {\bibfnamefont {V.~V.}\
  \bibnamefont {Naletov}}, \bibinfo {author} {\bibfnamefont {N.}~\bibnamefont
  {Beaulieu}}, \bibinfo {author} {\bibfnamefont {J.~B.}\ \bibnamefont
  {Youssef}}, \bibinfo {author} {\bibfnamefont {G.}~\bibnamefont {de~Loubens}},
  \bibinfo {author} {\bibfnamefont {Y.}~\bibnamefont {Tserkovnyak}},\ and\
  \bibinfo {author} {\bibfnamefont {O.}~\bibnamefont {Klein}},\ }\bibfield
  {title} {\bibinfo {title} {{Short-range thermal magnon diffusion in magnetic
  garnet}},\ }\href {https://doi.org/10.1103/PhysRevB.103.174432} {\bibfield
  {journal} {\bibinfo  {journal} {Physical Review B}\ }\textbf {\bibinfo
  {volume} {103}},\ \bibinfo {pages} {174432} (\bibinfo {year}
  {2021})}\BibitemShut {NoStop}%
\bibitem [{\citenamefont {Marković}\ \emph {et~al.}(2020)\citenamefont
  {Marković}, \citenamefont {Mizrahi}, \citenamefont {Querlioz},\ and\
  \citenamefont {Grollier}}]{Markovic2020}%
  \BibitemOpen
  \bibfield  {author} {\bibinfo {author} {\bibfnamefont {D.}~\bibnamefont
  {Marković}}, \bibinfo {author} {\bibfnamefont {A.}~\bibnamefont {Mizrahi}},
  \bibinfo {author} {\bibfnamefont {D.}~\bibnamefont {Querlioz}},\ and\
  \bibinfo {author} {\bibfnamefont {J.}~\bibnamefont {Grollier}},\ }\bibfield
  {title} {\bibinfo {title} {Physics for neuromorphic computing},\ }\href@noop
  {} {\bibfield  {journal} {\bibinfo  {journal} {Nature Reviews Physics}\
  }\textbf {\bibinfo {volume} {2}},\ \bibinfo {pages} {499} (\bibinfo {year}
  {2020})}\BibitemShut {NoStop}%
\bibitem [{\citenamefont {Kish}\ and\ \citenamefont
  {Granqvist}(2012)}]{Kish2012}%
  \BibitemOpen
  \bibfield  {author} {\bibinfo {author} {\bibfnamefont {L.~B.}\ \bibnamefont
  {Kish}}\ and\ \bibinfo {author} {\bibfnamefont {C.-G.}\ \bibnamefont
  {Granqvist}},\ }\bibfield  {title} {\bibinfo {title} {Electrical maxwell
  demon and szilard engine utilizing johnson noise, measurement, logic and
  control},\ }\href {https://doi.org/10.1371/journal.pone.0046800} {\bibfield
  {journal} {\bibinfo  {journal} {PLOS ONE}\ }\textbf {\bibinfo {volume} {7}},\
  \bibinfo {pages} {1} (\bibinfo {year} {2012})}\BibitemShut {NoStop}%
\bibitem [{\citenamefont {Koski}\ and\ \citenamefont
  {Pekola}(2016)}]{Koski2016}%
  \BibitemOpen
  \bibfield  {author} {\bibinfo {author} {\bibfnamefont {J.~V.}\ \bibnamefont
  {Koski}}\ and\ \bibinfo {author} {\bibfnamefont {J.~P.}\ \bibnamefont
  {Pekola}},\ }\bibfield  {title} {\bibinfo {title} {Maxwell's demons realized
  in electronic circuits},\ }\href
  {https://doi.org/https://doi.org/10.1016/j.crhy.2016.08.011} {\bibfield
  {journal} {\bibinfo  {journal} {Comptes Rendus Physique}\ }\textbf {\bibinfo
  {volume} {17}},\ \bibinfo {pages} {1130} (\bibinfo {year} {2016})},\ \bibinfo
  {note} {mesoscopic thermoelectric phenomena / Phénomènes thermoélectriques
  mésoscopiques}\BibitemShut {NoStop}%
\bibitem [{\citenamefont {Bergeret}\ \emph {et~al.}(2018)\citenamefont
  {Bergeret}, \citenamefont {Silaev}, \citenamefont {Virtanen},\ and\
  \citenamefont {Heikkil{\"{a}}}}]{Bergeret2018}%
  \BibitemOpen
  \bibfield  {author} {\bibinfo {author} {\bibfnamefont {F.~S.}\ \bibnamefont
  {Bergeret}}, \bibinfo {author} {\bibfnamefont {M.}~\bibnamefont {Silaev}},
  \bibinfo {author} {\bibfnamefont {P.}~\bibnamefont {Virtanen}},\ and\
  \bibinfo {author} {\bibfnamefont {T.~T.}\ \bibnamefont {Heikkil{\"{a}}}},\
  }\bibfield  {title} {\bibinfo {title} {{Colloquium: Nonequilibrium effects in
  superconductors with a spin-splitting field}},\ }\href
  {https://doi.org/10.1103/RevModPhys.90.041001} {\bibfield  {journal}
  {\bibinfo  {journal} {Reviews of Modern Physics}\ }\textbf {\bibinfo {volume}
  {90}},\ \bibinfo {pages} {041001} (\bibinfo {year} {2018})},\ \Eprint
  {https://arxiv.org/abs/1706.08245} {arXiv:1706.08245} \BibitemShut {NoStop}%
\bibitem [{\citenamefont {Heikkil{\"{a}}}\ \emph {et~al.}(2019)\citenamefont
  {Heikkil{\"{a}}}, \citenamefont {Silaev}, \citenamefont {Virtanen},\ and\
  \citenamefont {Bergeret}}]{Heikkila2019}%
  \BibitemOpen
  \bibfield  {author} {\bibinfo {author} {\bibfnamefont {T.~T.}\ \bibnamefont
  {Heikkil{\"{a}}}}, \bibinfo {author} {\bibfnamefont {M.}~\bibnamefont
  {Silaev}}, \bibinfo {author} {\bibfnamefont {P.}~\bibnamefont {Virtanen}},\
  and\ \bibinfo {author} {\bibfnamefont {F.~S.}\ \bibnamefont {Bergeret}},\
  }\bibfield  {title} {\bibinfo {title} {{Thermal, electric and spin transport
  in superconductor/ferromagnetic-insulator structures}},\ }\href
  {https://doi.org/10.1016/J.PROGSURF.2019.100540} {\bibfield  {journal}
  {\bibinfo  {journal} {Progress in Surface Science}\ }\textbf {\bibinfo
  {volume} {94}},\ \bibinfo {pages} {100540} (\bibinfo {year}
  {2019})}\BibitemShut {NoStop}%
\bibitem [{\citenamefont {Yang}\ \emph {et~al.}(2021)\citenamefont {Yang},
  \citenamefont {Ciccarelli},\ and\ \citenamefont {Robinson}}]{Yang2021}%
  \BibitemOpen
  \bibfield  {author} {\bibinfo {author} {\bibfnamefont {G.}~\bibnamefont
  {Yang}}, \bibinfo {author} {\bibfnamefont {C.}~\bibnamefont {Ciccarelli}},\
  and\ \bibinfo {author} {\bibfnamefont {J.~W.~A.}\ \bibnamefont {Robinson}},\
  }\bibfield  {title} {\bibinfo {title} {{Boosting spintronics with
  superconductivity}},\ }\href {https://doi.org/10.1063/5.0048904} {\bibfield
  {journal} {\bibinfo  {journal} {APL Materials}\ }\textbf {\bibinfo {volume}
  {9}},\ \bibinfo {pages} {050703} (\bibinfo {year} {2021})}\BibitemShut
  {NoStop}%
\bibitem [{\citenamefont {Jeon}\ \emph
  {et~al.}(2020{\natexlab{a}})\citenamefont {Jeon}, \citenamefont {Jeon},
  \citenamefont {Zhou}, \citenamefont {Migliorini}, \citenamefont {Yoon},\ and\
  \citenamefont {Parkin}}]{Jeon2020}%
  \BibitemOpen
  \bibfield  {author} {\bibinfo {author} {\bibfnamefont {K.-R.}\ \bibnamefont
  {Jeon}}, \bibinfo {author} {\bibfnamefont {J.-C.}\ \bibnamefont {Jeon}},
  \bibinfo {author} {\bibfnamefont {X.}~\bibnamefont {Zhou}}, \bibinfo {author}
  {\bibfnamefont {A.}~\bibnamefont {Migliorini}}, \bibinfo {author}
  {\bibfnamefont {J.}~\bibnamefont {Yoon}},\ and\ \bibinfo {author}
  {\bibfnamefont {S.~S.~P.}\ \bibnamefont {Parkin}},\ }\bibfield  {title}
  {\bibinfo {title} {{Giant Transition-State Quasiparticle Spin-Hall Effect in
  an Exchange-Spin-Split Superconductor Detected by Nonlocal Magnon Spin
  Transport}},\ }\href {https://doi.org/10.1021/ACSNANO.0C07187} {\bibfield
  {journal} {\bibinfo  {journal} {ACS Nano}\ }\textbf {\bibinfo {volume}
  {14}},\ \bibinfo {pages} {15874} (\bibinfo {year}
  {2020}{\natexlab{a}})}\BibitemShut {NoStop}%
\bibitem [{\citenamefont {Strambini}\ \emph {et~al.}(2017)\citenamefont
  {Strambini}, \citenamefont {Golovach}, \citenamefont {Simoni}, \citenamefont
  {Moodera}, \citenamefont {Bergeret},\ and\ \citenamefont
  {Giazotto}}]{Strambini2017}%
  \BibitemOpen
  \bibfield  {author} {\bibinfo {author} {\bibfnamefont {E.}~\bibnamefont
  {Strambini}}, \bibinfo {author} {\bibfnamefont {V.~N.}\ \bibnamefont
  {Golovach}}, \bibinfo {author} {\bibfnamefont {G.~D.}\ \bibnamefont
  {Simoni}}, \bibinfo {author} {\bibfnamefont {J.~S.}\ \bibnamefont {Moodera}},
  \bibinfo {author} {\bibfnamefont {F.~S.}\ \bibnamefont {Bergeret}},\ and\
  \bibinfo {author} {\bibfnamefont {F.}~\bibnamefont {Giazotto}},\ }\bibfield
  {title} {\bibinfo {title} {{Revealing the magnetic proximity effect in EuS/Al
  bilayers through superconducting tunneling spectroscopy}},\ }\href
  {https://doi.org/10.1103/PhysRevMaterials.1.054402} {\bibfield  {journal}
  {\bibinfo  {journal} {Physical Review Materials}\ }\textbf {\bibinfo {volume}
  {1}},\ \bibinfo {pages} {054402} (\bibinfo {year} {2017})}\BibitemShut
  {NoStop}%
\bibitem [{\citenamefont {Machon}\ \emph {et~al.}(2013)\citenamefont {Machon},
  \citenamefont {Eschrig},\ and\ \citenamefont {Belzig}}]{Machon2013}%
  \BibitemOpen
  \bibfield  {author} {\bibinfo {author} {\bibfnamefont {P.}~\bibnamefont
  {Machon}}, \bibinfo {author} {\bibfnamefont {M.}~\bibnamefont {Eschrig}},\
  and\ \bibinfo {author} {\bibfnamefont {W.}~\bibnamefont {Belzig}},\
  }\bibfield  {title} {\bibinfo {title} {{Nonlocal Thermoelectric Effects and
  Nonlocal Onsager relations in a Three-Terminal Proximity-Coupled
  Superconductor-Ferromagnet Device}},\ }\href
  {http://link.aps.org/doi/10.1103/PhysRevLett.110.047002} {\bibfield
  {journal} {\bibinfo  {journal} {Physical Review Letters}\ }\textbf {\bibinfo
  {volume} {110}},\ \bibinfo {pages} {47002} (\bibinfo {year}
  {2013})}\BibitemShut {NoStop}%
\bibitem [{\citenamefont {Machon}\ \emph {et~al.}(2014)\citenamefont {Machon},
  \citenamefont {Eschrig},\ and\ \citenamefont {Belzig}}]{Machon2014}%
  \BibitemOpen
  \bibfield  {author} {\bibinfo {author} {\bibfnamefont {P.}~\bibnamefont
  {Machon}}, \bibinfo {author} {\bibfnamefont {M.}~\bibnamefont {Eschrig}},\
  and\ \bibinfo {author} {\bibfnamefont {W.}~\bibnamefont {Belzig}},\
  }\bibfield  {title} {\bibinfo {title} {{Giant thermoelectric effects in a
  proximity-coupled superconductor–ferromagnet device}},\ }\href
  {http://stacks.iop.org/1367-2630/16/i=7/a=073002} {\bibfield  {journal}
  {\bibinfo  {journal} {New Journal of Physics}\ }\textbf {\bibinfo {volume}
  {16}},\ \bibinfo {pages} {73002} (\bibinfo {year} {2014})}\BibitemShut
  {NoStop}%
\bibitem [{\citenamefont {Ozaeta}\ \emph {et~al.}(2014)\citenamefont {Ozaeta},
  \citenamefont {Virtanen}, \citenamefont {Bergeret},\ and\ \citenamefont
  {Heikkil{\"{a}}}}]{Ozaeta2014}%
  \BibitemOpen
  \bibfield  {author} {\bibinfo {author} {\bibfnamefont {A.}~\bibnamefont
  {Ozaeta}}, \bibinfo {author} {\bibfnamefont {P.}~\bibnamefont {Virtanen}},
  \bibinfo {author} {\bibfnamefont {F.~S.}\ \bibnamefont {Bergeret}},\ and\
  \bibinfo {author} {\bibfnamefont {T.~T.}\ \bibnamefont {Heikkil{\"{a}}}},\
  }\bibfield  {title} {\bibinfo {title} {{Predicted very large thermoelectric
  effect in ferromagnet-superconductor junctions in the presence of a
  spin-splitting magnetic field}},\ }\href
  {https://doi.org/10.1103/PhysRevLett.112.057001} {\bibfield  {journal}
  {\bibinfo  {journal} {Physical Review Letters}\ }\textbf {\bibinfo {volume}
  {112}},\ \bibinfo {pages} {057001} (\bibinfo {year} {2014})}\BibitemShut
  {NoStop}%
\bibitem [{\citenamefont {Wolf}\ \emph {et~al.}(2014)\citenamefont {Wolf},
  \citenamefont {S{\"{u}}rgers}, \citenamefont {Fischer},\ and\ \citenamefont
  {Beckmann}}]{Wolf2014}%
  \BibitemOpen
  \bibfield  {author} {\bibinfo {author} {\bibfnamefont {M.~J.}\ \bibnamefont
  {Wolf}}, \bibinfo {author} {\bibfnamefont {C.}~\bibnamefont {S{\"{u}}rgers}},
  \bibinfo {author} {\bibfnamefont {G.}~\bibnamefont {Fischer}},\ and\ \bibinfo
  {author} {\bibfnamefont {D.}~\bibnamefont {Beckmann}},\ }\bibfield  {title}
  {\bibinfo {title} {{Spin-polarized quasiparticle transport in exchange-split
  superconducting aluminum on europium sulfide}},\ }\href
  {https://doi.org/10.1103/PhysRevB.90.144509} {\bibfield  {journal} {\bibinfo
  {journal} {Physical Review B}\ }\textbf {\bibinfo {volume} {90}},\ \bibinfo
  {pages} {144509} (\bibinfo {year} {2014})}\BibitemShut {NoStop}%
\bibitem [{\citenamefont {Golovchanskiy}\ \emph {et~al.}(2018)\citenamefont
  {Golovchanskiy}, \citenamefont {Abramov}, \citenamefont {Stolyarov},
  \citenamefont {Bolginov}, \citenamefont {Ryazanov}, \citenamefont {Golubov},\
  and\ \citenamefont {Ustinov}}]{Golovchanskiy2018}%
  \BibitemOpen
  \bibfield  {author} {\bibinfo {author} {\bibfnamefont {I.~A.}\ \bibnamefont
  {Golovchanskiy}}, \bibinfo {author} {\bibfnamefont {N.~N.}\ \bibnamefont
  {Abramov}}, \bibinfo {author} {\bibfnamefont {V.~S.}\ \bibnamefont
  {Stolyarov}}, \bibinfo {author} {\bibfnamefont {V.~V.}\ \bibnamefont
  {Bolginov}}, \bibinfo {author} {\bibfnamefont {V.~V.}\ \bibnamefont
  {Ryazanov}}, \bibinfo {author} {\bibfnamefont {A.~A.}\ \bibnamefont
  {Golubov}},\ and\ \bibinfo {author} {\bibfnamefont {A.~V.}\ \bibnamefont
  {Ustinov}},\ }\bibfield  {title} {\bibinfo {title}
  {{Ferromagnet/Superconductor Hybridization for Magnonic Applications}},\
  }\href {https://doi.org/10.1002/ADFM.201802375} {\bibfield  {journal}
  {\bibinfo  {journal} {Advanced Functional Materials}\ }\textbf {\bibinfo
  {volume} {28}},\ \bibinfo {pages} {1802375} (\bibinfo {year}
  {2018})}\BibitemShut {NoStop}%
\bibitem [{\citenamefont {Dobrovolskiy}\ \emph {et~al.}(2019)\citenamefont
  {Dobrovolskiy}, \citenamefont {Sachser}, \citenamefont {Br{\"{a}}cher},
  \citenamefont {B{\"{o}}ttcher}, \citenamefont {Kruglyak}, \citenamefont
  {Vovk}, \citenamefont {Shklovskij}, \citenamefont {Huth}, \citenamefont
  {Hillebrands},\ and\ \citenamefont {Chumak}}]{Dobrovolskiy2019}%
  \BibitemOpen
  \bibfield  {author} {\bibinfo {author} {\bibfnamefont {O.~V.}\ \bibnamefont
  {Dobrovolskiy}}, \bibinfo {author} {\bibfnamefont {R.}~\bibnamefont
  {Sachser}}, \bibinfo {author} {\bibfnamefont {T.}~\bibnamefont
  {Br{\"{a}}cher}}, \bibinfo {author} {\bibfnamefont {T.}~\bibnamefont
  {B{\"{o}}ttcher}}, \bibinfo {author} {\bibfnamefont {V.~V.}\ \bibnamefont
  {Kruglyak}}, \bibinfo {author} {\bibfnamefont {R.~V.}\ \bibnamefont {Vovk}},
  \bibinfo {author} {\bibfnamefont {V.~A.}\ \bibnamefont {Shklovskij}},
  \bibinfo {author} {\bibfnamefont {M.}~\bibnamefont {Huth}}, \bibinfo {author}
  {\bibfnamefont {B.}~\bibnamefont {Hillebrands}},\ and\ \bibinfo {author}
  {\bibfnamefont {A.~V.}\ \bibnamefont {Chumak}},\ }\bibfield  {title}
  {\bibinfo {title} {{Magnon–fluxon interaction in a
  ferromagnet/superconductor heterostructure}},\ }\href
  {https://doi.org/10.1038/s41567-019-0428-5} {\bibfield  {journal} {\bibinfo
  {journal} {Nature Physics 2019 15:5}\ }\textbf {\bibinfo {volume} {15}},\
  \bibinfo {pages} {477} (\bibinfo {year} {2019})}\BibitemShut {NoStop}%
\bibitem [{\citenamefont {Li}\ \emph {et~al.}(2019)\citenamefont {Li},
  \citenamefont {Polakovic}, \citenamefont {Wang}, \citenamefont {Xu},
  \citenamefont {Lendinez}, \citenamefont {Zhang}, \citenamefont {Ding},
  \citenamefont {Khaire}, \citenamefont {Saglam}, \citenamefont {Divan},
  \citenamefont {Pearson}, \citenamefont {Kwok}, \citenamefont {Xiao},
  \citenamefont {Novosad}, \citenamefont {Hoffmann},\ and\ \citenamefont
  {Zhang}}]{Li2019}%
  \BibitemOpen
  \bibfield  {author} {\bibinfo {author} {\bibfnamefont {Y.}~\bibnamefont
  {Li}}, \bibinfo {author} {\bibfnamefont {T.}~\bibnamefont {Polakovic}},
  \bibinfo {author} {\bibfnamefont {Y.-L.}\ \bibnamefont {Wang}}, \bibinfo
  {author} {\bibfnamefont {J.}~\bibnamefont {Xu}}, \bibinfo {author}
  {\bibfnamefont {S.}~\bibnamefont {Lendinez}}, \bibinfo {author}
  {\bibfnamefont {Z.}~\bibnamefont {Zhang}}, \bibinfo {author} {\bibfnamefont
  {J.}~\bibnamefont {Ding}}, \bibinfo {author} {\bibfnamefont {T.}~\bibnamefont
  {Khaire}}, \bibinfo {author} {\bibfnamefont {H.}~\bibnamefont {Saglam}},
  \bibinfo {author} {\bibfnamefont {R.}~\bibnamefont {Divan}}, \bibinfo
  {author} {\bibfnamefont {J.}~\bibnamefont {Pearson}}, \bibinfo {author}
  {\bibfnamefont {W.-K.}\ \bibnamefont {Kwok}}, \bibinfo {author}
  {\bibfnamefont {Z.}~\bibnamefont {Xiao}}, \bibinfo {author} {\bibfnamefont
  {V.}~\bibnamefont {Novosad}}, \bibinfo {author} {\bibfnamefont
  {A.}~\bibnamefont {Hoffmann}},\ and\ \bibinfo {author} {\bibfnamefont
  {W.}~\bibnamefont {Zhang}},\ }\bibfield  {title} {\bibinfo {title} {{Strong
  Coupling between Magnons and Microwave Photons in On-Chip
  Ferromagnet-Superconductor Thin-Film Devices}},\ }\href
  {https://doi.org/10.1103/PhysRevLett.123.107701} {\bibfield  {journal}
  {\bibinfo  {journal} {Physical Review Letters}\ }\textbf {\bibinfo {volume}
  {123}},\ \bibinfo {pages} {107701} (\bibinfo {year} {2019})}\BibitemShut
  {NoStop}%
\bibitem [{\citenamefont {Jeon}\ \emph {et~al.}(2019)\citenamefont {Jeon},
  \citenamefont {Ciccarelli}, \citenamefont {Kurebayashi}, \citenamefont
  {Cohen}, \citenamefont {Montiel}, \citenamefont {Eschrig}, \citenamefont
  {Komori}, \citenamefont {Robinson},\ and\ \citenamefont
  {Blamire}}]{Jeon2019}%
  \BibitemOpen
  \bibfield  {author} {\bibinfo {author} {\bibfnamefont {K.~R.}\ \bibnamefont
  {Jeon}}, \bibinfo {author} {\bibfnamefont {C.}~\bibnamefont {Ciccarelli}},
  \bibinfo {author} {\bibfnamefont {H.}~\bibnamefont {Kurebayashi}}, \bibinfo
  {author} {\bibfnamefont {L.~F.}\ \bibnamefont {Cohen}}, \bibinfo {author}
  {\bibfnamefont {X.}~\bibnamefont {Montiel}}, \bibinfo {author} {\bibfnamefont
  {M.}~\bibnamefont {Eschrig}}, \bibinfo {author} {\bibfnamefont
  {S.}~\bibnamefont {Komori}}, \bibinfo {author} {\bibfnamefont {J.~W.}\
  \bibnamefont {Robinson}},\ and\ \bibinfo {author} {\bibfnamefont {M.~G.}\
  \bibnamefont {Blamire}},\ }\bibfield  {title} {\bibinfo {title}
  {{Exchange-field enhancement of superconducting spin pumping}},\ }\href
  {https://doi.org/10.1103/PhysRevB.99.024507} {\bibfield  {journal} {\bibinfo
  {journal} {Physical Review B}\ }\textbf {\bibinfo {volume} {99}},\ \bibinfo
  {pages} {024507} (\bibinfo {year} {2019})}\BibitemShut {NoStop}%
\bibitem [{\citenamefont {Lachance-Quirion}\ \emph {et~al.}(2020)\citenamefont
  {Lachance-Quirion}, \citenamefont {Wolski}, \citenamefont {Tabuchi},
  \citenamefont {Kono}, \citenamefont {Usami},\ and\ \citenamefont
  {Nakamura}}]{Lachance-Quirion2020}%
  \BibitemOpen
  \bibfield  {author} {\bibinfo {author} {\bibfnamefont {D.}~\bibnamefont
  {Lachance-Quirion}}, \bibinfo {author} {\bibfnamefont {S.~P.}\ \bibnamefont
  {Wolski}}, \bibinfo {author} {\bibfnamefont {Y.}~\bibnamefont {Tabuchi}},
  \bibinfo {author} {\bibfnamefont {S.}~\bibnamefont {Kono}}, \bibinfo {author}
  {\bibfnamefont {K.}~\bibnamefont {Usami}},\ and\ \bibinfo {author}
  {\bibfnamefont {Y.}~\bibnamefont {Nakamura}},\ }\bibfield  {title} {\bibinfo
  {title} {{Entanglement-based single-shot detection of a single magnon with a
  superconducting qubit}},\ }\href {https://doi.org/10.1126/SCIENCE.AAZ9236}
  {\bibfield  {journal} {\bibinfo  {journal} {Science}\ }\textbf {\bibinfo
  {volume} {367}},\ \bibinfo {pages} {425} (\bibinfo {year}
  {2020})}\BibitemShut {NoStop}%
\bibitem [{\citenamefont {Jeon}\ \emph
  {et~al.}(2020{\natexlab{b}})\citenamefont {Jeon}, \citenamefont {Montiel},
  \citenamefont {Komori}, \citenamefont {Ciccarelli}, \citenamefont {Haigh},
  \citenamefont {Kurebayashi}, \citenamefont {Cohen}, \citenamefont {Chan},
  \citenamefont {Stenning}, \citenamefont {Lee}, \citenamefont {Eschrig},
  \citenamefont {Blamire},\ and\ \citenamefont {Robinson}}]{Jeon2020a}%
  \BibitemOpen
  \bibfield  {author} {\bibinfo {author} {\bibfnamefont {K.-R.}\ \bibnamefont
  {Jeon}}, \bibinfo {author} {\bibfnamefont {X.}~\bibnamefont {Montiel}},
  \bibinfo {author} {\bibfnamefont {S.}~\bibnamefont {Komori}}, \bibinfo
  {author} {\bibfnamefont {C.}~\bibnamefont {Ciccarelli}}, \bibinfo {author}
  {\bibfnamefont {J.}~\bibnamefont {Haigh}}, \bibinfo {author} {\bibfnamefont
  {H.}~\bibnamefont {Kurebayashi}}, \bibinfo {author} {\bibfnamefont {L.~F.}\
  \bibnamefont {Cohen}}, \bibinfo {author} {\bibfnamefont {A.~K.}\ \bibnamefont
  {Chan}}, \bibinfo {author} {\bibfnamefont {K.~D.}\ \bibnamefont {Stenning}},
  \bibinfo {author} {\bibfnamefont {C.-M.}\ \bibnamefont {Lee}}, \bibinfo
  {author} {\bibfnamefont {M.}~\bibnamefont {Eschrig}}, \bibinfo {author}
  {\bibfnamefont {M.~G.}\ \bibnamefont {Blamire}},\ and\ \bibinfo {author}
  {\bibfnamefont {J.~W.}\ \bibnamefont {Robinson}},\ }\bibfield  {title}
  {\bibinfo {title} {{Tunable Pure Spin Supercurrents and the Demonstration of
  Their Gateability in a Spin-Wave Device}},\ }\href
  {https://doi.org/10.1103/PhysRevX.10.031020} {\bibfield  {journal} {\bibinfo
  {journal} {Physical Review X}\ }\textbf {\bibinfo {volume} {10}},\ \bibinfo
  {pages} {031020} (\bibinfo {year} {2020}{\natexlab{b}})}\BibitemShut
  {NoStop}%
\bibitem [{\citenamefont {Golovchanskiy}\ \emph {et~al.}(2020)\citenamefont
  {Golovchanskiy}, \citenamefont {Abramov}, \citenamefont {Stolyarov},
  \citenamefont {Chichkov}, \citenamefont {Silaev}, \citenamefont {Shchetinin},
  \citenamefont {Golubov}, \citenamefont {Ryazanov}, \citenamefont {Ustinov},\
  and\ \citenamefont {Kupriyanov}}]{Golovchanskiy2020}%
  \BibitemOpen
  \bibfield  {author} {\bibinfo {author} {\bibfnamefont {I.}~\bibnamefont
  {Golovchanskiy}}, \bibinfo {author} {\bibfnamefont {N.}~\bibnamefont
  {Abramov}}, \bibinfo {author} {\bibfnamefont {V.}~\bibnamefont {Stolyarov}},
  \bibinfo {author} {\bibfnamefont {V.}~\bibnamefont {Chichkov}}, \bibinfo
  {author} {\bibfnamefont {M.}~\bibnamefont {Silaev}}, \bibinfo {author}
  {\bibfnamefont {I.}~\bibnamefont {Shchetinin}}, \bibinfo {author}
  {\bibfnamefont {A.}~\bibnamefont {Golubov}}, \bibinfo {author} {\bibfnamefont
  {V.}~\bibnamefont {Ryazanov}}, \bibinfo {author} {\bibfnamefont
  {A.}~\bibnamefont {Ustinov}},\ and\ \bibinfo {author} {\bibfnamefont
  {M.}~\bibnamefont {Kupriyanov}},\ }\bibfield  {title} {\bibinfo {title}
  {Magnetization dynamics in proximity-coupled
  superconductor-ferromagnet-superconductor multilayers},\ }\href
  {https://doi.org/10.1103/PhysRevApplied.14.024086} {\bibfield  {journal}
  {\bibinfo  {journal} {Phys. Rev. Applied}\ }\textbf {\bibinfo {volume}
  {14}},\ \bibinfo {pages} {024086} (\bibinfo {year} {2020})}\BibitemShut
  {NoStop}%
\bibitem [{\citenamefont {Golovchanskiy}\ \emph {et~al.}(2021)\citenamefont
  {Golovchanskiy}, \citenamefont {Abramov}, \citenamefont {Stolyarov},
  \citenamefont {Weides}, \citenamefont {Ryazanov}, \citenamefont {Golubov},
  \citenamefont {Ustinov},\ and\ \citenamefont
  {Kupriyanov}}]{Golovchanskiy2021}%
  \BibitemOpen
  \bibfield  {author} {\bibinfo {author} {\bibfnamefont {I.}~\bibnamefont
  {Golovchanskiy}}, \bibinfo {author} {\bibfnamefont {N.}~\bibnamefont
  {Abramov}}, \bibinfo {author} {\bibfnamefont {V.}~\bibnamefont {Stolyarov}},
  \bibinfo {author} {\bibfnamefont {M.}~\bibnamefont {Weides}}, \bibinfo
  {author} {\bibfnamefont {V.}~\bibnamefont {Ryazanov}}, \bibinfo {author}
  {\bibfnamefont {A.}~\bibnamefont {Golubov}}, \bibinfo {author} {\bibfnamefont
  {A.}~\bibnamefont {Ustinov}},\ and\ \bibinfo {author} {\bibfnamefont
  {M.}~\bibnamefont {Kupriyanov}},\ }\bibfield  {title} {\bibinfo {title}
  {Ultrastrong photon-to-magnon coupling in multilayered heterostructures
  involving superconducting coherence via ferromagnetic layers},\ }\href
  {https://doi.org/10.1126/sciadv.abe8638} {\bibfield  {journal} {\bibinfo
  {journal} {Science Advances}\ }\textbf {\bibinfo {volume} {7}},\ \bibinfo
  {pages} {eabe8638} (\bibinfo {year} {2021})}\BibitemShut {NoStop}%
\bibitem [{\citenamefont {Ohnuma}\ \emph {et~al.}(2017)\citenamefont {Ohnuma},
  \citenamefont {Matsuo},\ and\ \citenamefont {Maekawa}}]{Ohnuma2017}%
  \BibitemOpen
  \bibfield  {author} {\bibinfo {author} {\bibfnamefont {Y.}~\bibnamefont
  {Ohnuma}}, \bibinfo {author} {\bibfnamefont {M.}~\bibnamefont {Matsuo}},\
  and\ \bibinfo {author} {\bibfnamefont {S.}~\bibnamefont {Maekawa}},\
  }\bibfield  {title} {\bibinfo {title} {{Theory of the spin Peltier effect}},\
  }\href {https://doi.org/10.1103/PhysRevB.96.134412} {\bibfield  {journal}
  {\bibinfo  {journal} {Physical Review B}\ }\textbf {\bibinfo {volume} {96}},\
  \bibinfo {pages} {134412} (\bibinfo {year} {2017})}\BibitemShut {NoStop}%
\bibitem [{\citenamefont {Kato}\ \emph {et~al.}(2019)\citenamefont {Kato},
  \citenamefont {Ohnuma}, \citenamefont {Matsuo}, \citenamefont {Rech},
  \citenamefont {Jonckheere},\ and\ \citenamefont {Martin}}]{Kato2019}%
  \BibitemOpen
  \bibfield  {author} {\bibinfo {author} {\bibfnamefont {T.}~\bibnamefont
  {Kato}}, \bibinfo {author} {\bibfnamefont {Y.}~\bibnamefont {Ohnuma}},
  \bibinfo {author} {\bibfnamefont {M.}~\bibnamefont {Matsuo}}, \bibinfo
  {author} {\bibfnamefont {J.}~\bibnamefont {Rech}}, \bibinfo {author}
  {\bibfnamefont {T.}~\bibnamefont {Jonckheere}},\ and\ \bibinfo {author}
  {\bibfnamefont {T.}~\bibnamefont {Martin}},\ }\bibfield  {title} {\bibinfo
  {title} {{Microscopic theory of spin transport at the interface between a
  superconductor and a ferromagnetic insulator}},\ }\href
  {https://doi.org/10.1103/PhysRevB.99.144411} {\bibfield  {journal} {\bibinfo
  {journal} {Physical Review B}\ }\textbf {\bibinfo {volume} {99}},\ \bibinfo
  {pages} {144411} (\bibinfo {year} {2019})}\BibitemShut {NoStop}%
\bibitem [{\citenamefont {Chakraborty}\ and\ \citenamefont
  {Heikkil{\"{a}}}(2019)}]{Chakraborty2019}%
  \BibitemOpen
  \bibfield  {author} {\bibinfo {author} {\bibfnamefont {S.}~\bibnamefont
  {Chakraborty}}\ and\ \bibinfo {author} {\bibfnamefont {T.~T.}\ \bibnamefont
  {Heikkil{\"{a}}}},\ }\bibfield  {title} {\bibinfo {title} {{Thermalization of
  hot electrons via interfacial electron-magnon interaction}},\ }\href
  {https://doi.org/10.1103/PhysRevB.100.035423} {\bibfield  {journal} {\bibinfo
   {journal} {Physical Review B}\ }\textbf {\bibinfo {volume} {100}},\ \bibinfo
  {pages} {035423} (\bibinfo {year} {2019})}\BibitemShut {NoStop}%
\bibitem [{\citenamefont {Vargas}\ and\ \citenamefont
  {Moura}(2020{\natexlab{a}})}]{Vargas2020}%
  \BibitemOpen
  \bibfield  {author} {\bibinfo {author} {\bibfnamefont {V.~S. U.~A.}\
  \bibnamefont {Vargas}}\ and\ \bibinfo {author} {\bibfnamefont {A.~R.}\
  \bibnamefont {Moura}},\ }\bibfield  {title} {\bibinfo {title} {{Spin current
  injection at magnetic insulator/superconductor interfaces}},\ }\href
  {https://doi.org/10.1103/PhysRevB.102.024412} {\bibfield  {journal} {\bibinfo
   {journal} {Physical Review B}\ }\textbf {\bibinfo {volume} {102}},\ \bibinfo
  {pages} {024412} (\bibinfo {year} {2020}{\natexlab{a}})}\BibitemShut
  {NoStop}%
\bibitem [{\citenamefont {Vargas}\ and\ \citenamefont
  {Moura}(2020{\natexlab{b}})}]{Vargas2020a}%
  \BibitemOpen
  \bibfield  {author} {\bibinfo {author} {\bibfnamefont {V.~S.}\ \bibnamefont
  {Vargas}}\ and\ \bibinfo {author} {\bibfnamefont {A.~R.}\ \bibnamefont
  {Moura}},\ }\bibfield  {title} {\bibinfo {title} {{Injection of spin current
  at the superconductor/ferromagnetic insulator interface}},\ }\href
  {https://doi.org/10.1016/J.JMMM.2019.165813} {\bibfield  {journal} {\bibinfo
  {journal} {Journal of Magnetism and Magnetic Materials}\ }\textbf {\bibinfo
  {volume} {494}},\ \bibinfo {pages} {165813} (\bibinfo {year}
  {2020}{\natexlab{b}})}\BibitemShut {NoStop}%
\bibitem [{\citenamefont {Ojaj{\"{a}}rvi}\ \emph {et~al.}(2021)\citenamefont
  {Ojaj{\"{a}}rvi}, \citenamefont {Heikkil{\"{a}}}, \citenamefont {Virtanen},\
  and\ \citenamefont {Silaev}}]{Ojajarvi2021}%
  \BibitemOpen
  \bibfield  {author} {\bibinfo {author} {\bibfnamefont {R.}~\bibnamefont
  {Ojaj{\"{a}}rvi}}, \bibinfo {author} {\bibfnamefont {T.~T.}\ \bibnamefont
  {Heikkil{\"{a}}}}, \bibinfo {author} {\bibfnamefont {P.}~\bibnamefont
  {Virtanen}},\ and\ \bibinfo {author} {\bibfnamefont {M.~A.}\ \bibnamefont
  {Silaev}},\ }\bibfield  {title} {\bibinfo {title} {{Giant enhancement to spin
  battery effect in superconductor/ferromagnetic insulator systems}},\ }\href
  {https://doi.org/10.1103/PhysRevB.103.224524} {\bibfield  {journal} {\bibinfo
   {journal} {Physical Review B}\ }\textbf {\bibinfo {volume} {103}},\ \bibinfo
  {pages} {224524} (\bibinfo {year} {2021})}\BibitemShut {NoStop}%
\bibitem [{\citenamefont {Ahari}\ and\ \citenamefont
  {Tserkovnyak}(2021)}]{Ahari2021}%
  \BibitemOpen
  \bibfield  {author} {\bibinfo {author} {\bibfnamefont {M.~T.}\ \bibnamefont
  {Ahari}}\ and\ \bibinfo {author} {\bibfnamefont {Y.}~\bibnamefont
  {Tserkovnyak}},\ }\bibfield  {title} {\bibinfo {title}
  {{Superconductivity-enhanced spin pumping: Role of Andreev resonances}},\
  }\href {https://doi.org/10.1103/PhysRevB.103.L100406} {\bibfield  {journal}
  {\bibinfo  {journal} {Physical Review B}\ }\textbf {\bibinfo {volume}
  {103}},\ \bibinfo {pages} {L100406} (\bibinfo {year} {2021})}\BibitemShut
  {NoStop}%
\bibitem [{\citenamefont {Bardeen}(1962)}]{Bardeen1962}%
  \BibitemOpen
  \bibfield  {author} {\bibinfo {author} {\bibfnamefont {J.}~\bibnamefont
  {Bardeen}},\ }\bibfield  {title} {\bibinfo {title} {Critical fields and
  currents in superconductors},\ }\href
  {https://doi.org/10.1103/RevModPhys.34.667} {\bibfield  {journal} {\bibinfo
  {journal} {Rev. Mod. Phys.}\ }\textbf {\bibinfo {volume} {34}},\ \bibinfo
  {pages} {667} (\bibinfo {year} {1962})}\BibitemShut {NoStop}%
\bibitem [{\citenamefont {Moraru}\ \emph {et~al.}(2006)\citenamefont {Moraru},
  \citenamefont {Pratt},\ and\ \citenamefont {Birge}}]{Moraru2006}%
  \BibitemOpen
  \bibfield  {author} {\bibinfo {author} {\bibfnamefont {I.~C.}\ \bibnamefont
  {Moraru}}, \bibinfo {author} {\bibfnamefont {W.~P.}\ \bibnamefont {Pratt}},\
  and\ \bibinfo {author} {\bibfnamefont {N.~O.}\ \bibnamefont {Birge}},\
  }\bibfield  {title} {\bibinfo {title} {Magnetization-dependent ${T}_{c}$
  shift in ferromagnet/superconductor/ferromagnet trilayers with a strong
  ferromagnet},\ }\href {https://doi.org/10.1103/PhysRevLett.96.037004}
  {\bibfield  {journal} {\bibinfo  {journal} {Phys. Rev. Lett.}\ }\textbf
  {\bibinfo {volume} {96}},\ \bibinfo {pages} {037004} (\bibinfo {year}
  {2006})}\BibitemShut {NoStop}%
\bibitem [{\citenamefont {Li}\ \emph {et~al.}(2013)\citenamefont {Li},
  \citenamefont {Roschewsky}, \citenamefont {Assaf}, \citenamefont {Eich},
  \citenamefont {Epstein-Martin}, \citenamefont {Heiman}, \citenamefont
  {M\"unzenberg},\ and\ \citenamefont {Moodera}}]{Li2013}%
  \BibitemOpen
  \bibfield  {author} {\bibinfo {author} {\bibfnamefont {B.}~\bibnamefont
  {Li}}, \bibinfo {author} {\bibfnamefont {N.}~\bibnamefont {Roschewsky}},
  \bibinfo {author} {\bibfnamefont {B.~A.}\ \bibnamefont {Assaf}}, \bibinfo
  {author} {\bibfnamefont {M.}~\bibnamefont {Eich}}, \bibinfo {author}
  {\bibfnamefont {M.}~\bibnamefont {Epstein-Martin}}, \bibinfo {author}
  {\bibfnamefont {D.}~\bibnamefont {Heiman}}, \bibinfo {author} {\bibfnamefont
  {M.}~\bibnamefont {M\"unzenberg}},\ and\ \bibinfo {author} {\bibfnamefont
  {J.~S.}\ \bibnamefont {Moodera}},\ }\bibfield  {title} {\bibinfo {title}
  {Superconducting spin switch with infinite magnetoresistance induced by an
  internal exchange field},\ }\href
  {https://doi.org/10.1103/PhysRevLett.110.097001} {\bibfield  {journal}
  {\bibinfo  {journal} {Phys. Rev. Lett.}\ }\textbf {\bibinfo {volume} {110}},\
  \bibinfo {pages} {097001} (\bibinfo {year} {2013})}\BibitemShut {NoStop}%
\bibitem [{\citenamefont {Nakata}\ and\ \citenamefont
  {Ohnuma}(2021)}]{Nakata2021}%
  \BibitemOpen
  \bibfield  {author} {\bibinfo {author} {\bibfnamefont {K.}~\bibnamefont
  {Nakata}}\ and\ \bibinfo {author} {\bibfnamefont {Y.}~\bibnamefont
  {Ohnuma}},\ }\bibfield  {title} {\bibinfo {title} {Magnonic thermal transport
  using the quantum boltzmann equation},\ }\href
  {https://doi.org/10.1103/PhysRevB.104.064408} {\bibfield  {journal} {\bibinfo
   {journal} {Phys. Rev. B}\ }\textbf {\bibinfo {volume} {104}},\ \bibinfo
  {pages} {064408} (\bibinfo {year} {2021})}\BibitemShut {NoStop}%
\bibitem [{\citenamefont {Taylor}\ \emph {et~al.}(2002)\citenamefont {Taylor},
  \citenamefont {Taylor},\ and\ \citenamefont {Heinonen}}]{Taylor2002}%
  \BibitemOpen
  \bibfield  {author} {\bibinfo {author} {\bibfnamefont {P.}~\bibnamefont
  {Taylor}}, \bibinfo {author} {\bibfnamefont {P.}~\bibnamefont {Taylor}},\
  and\ \bibinfo {author} {\bibfnamefont {O.}~\bibnamefont {Heinonen}},\ }\href
  {https://books.google.es/books?id=hyx6BjEX4U8C} {\emph {\bibinfo {title} {A
  Quantum Approach to Condensed Matter Physics}}}\ (\bibinfo  {publisher}
  {Cambridge University Press},\ \bibinfo {year} {2002})\BibitemShut {NoStop}%
\bibitem [{\citenamefont {Heikkil{\"a}}(2013)}]{Heikkila2013}%
  \BibitemOpen
  \bibfield  {author} {\bibinfo {author} {\bibfnamefont {T.}~\bibnamefont
  {Heikkil{\"a}}},\ }\href@noop {} {\emph {\bibinfo {title} {{The Physics of
  Nanoelectronics: Transport and Fluctuation Phenomena at Low
  Temperatures}}}},\ Oxford Master Series in Physics\ (\bibinfo  {publisher}
  {OUP Oxford},\ \bibinfo {year} {2013})\BibitemShut {NoStop}%
\bibitem [{\citenamefont {Giazotto}\ \emph {et~al.}(2006)\citenamefont
  {Giazotto}, \citenamefont {Heikkil\"a}, \citenamefont {Luukanen},
  \citenamefont {Savin},\ and\ \citenamefont {Pekola}}]{Giazotto2006}%
  \BibitemOpen
  \bibfield  {author} {\bibinfo {author} {\bibfnamefont {F.}~\bibnamefont
  {Giazotto}}, \bibinfo {author} {\bibfnamefont {T.~T.}\ \bibnamefont
  {Heikkil\"a}}, \bibinfo {author} {\bibfnamefont {A.}~\bibnamefont
  {Luukanen}}, \bibinfo {author} {\bibfnamefont {A.~M.}\ \bibnamefont
  {Savin}},\ and\ \bibinfo {author} {\bibfnamefont {J.~P.}\ \bibnamefont
  {Pekola}},\ }\bibfield  {title} {\bibinfo {title} {Opportunities for
  mesoscopics in thermometry and refrigeration: Physics and applications},\
  }\href {https://doi.org/10.1103/RevModPhys.78.217} {\bibfield  {journal}
  {\bibinfo  {journal} {Rev. Mod. Phys.}\ }\textbf {\bibinfo {volume} {78}},\
  \bibinfo {pages} {217} (\bibinfo {year} {2006})}\BibitemShut {NoStop}%
\bibitem [{\citenamefont {Kamra}\ \emph {et~al.}(2018)\citenamefont {Kamra},
  \citenamefont {Rezaei},\ and\ \citenamefont {Belzig}}]{Kamra2018}%
  \BibitemOpen
  \bibfield  {author} {\bibinfo {author} {\bibfnamefont {A.}~\bibnamefont
  {Kamra}}, \bibinfo {author} {\bibfnamefont {A.}~\bibnamefont {Rezaei}},\ and\
  \bibinfo {author} {\bibfnamefont {W.}~\bibnamefont {Belzig}},\ }\bibfield
  {title} {\bibinfo {title} {Spin splitting induced in a superconductor by an
  antiferromagnetic insulator},\ }\href
  {https://doi.org/10.1103/PhysRevLett.121.247702} {\bibfield  {journal}
  {\bibinfo  {journal} {Phys. Rev. Lett.}\ }\textbf {\bibinfo {volume} {121}},\
  \bibinfo {pages} {247702} (\bibinfo {year} {2018})}\BibitemShut {NoStop}%
\bibitem [{\citenamefont {Holstein}\ and\ \citenamefont
  {Primakoff}(1940)}]{Holstein1940}%
  \BibitemOpen
  \bibfield  {author} {\bibinfo {author} {\bibfnamefont {T.}~\bibnamefont
  {Holstein}}\ and\ \bibinfo {author} {\bibfnamefont {H.}~\bibnamefont
  {Primakoff}},\ }\bibfield  {title} {\bibinfo {title} {Field dependence of the
  intrinsic domain magnetization of a ferromagnet},\ }\href
  {https://doi.org/10.1103/PhysRev.58.1098} {\bibfield  {journal} {\bibinfo
  {journal} {Phys. Rev.}\ }\textbf {\bibinfo {volume} {58}},\ \bibinfo {pages}
  {1098} (\bibinfo {year} {1940})}\BibitemShut {NoStop}%
\bibitem [{\citenamefont {Bobkova}\ and\ \citenamefont
  {Bobkov}(2011)}]{Bobkova2011}%
  \BibitemOpen
  \bibfield  {author} {\bibinfo {author} {\bibfnamefont {I.~V.}\ \bibnamefont
  {Bobkova}}\ and\ \bibinfo {author} {\bibfnamefont {A.~M.}\ \bibnamefont
  {Bobkov}},\ }\bibfield  {title} {\bibinfo {title} {Recovering the
  superconducting state via spin accumulation above the pair-breaking magnetic
  field of superconductor/ferromagnet multilayers},\ }\href
  {https://doi.org/10.1103/PhysRevB.84.140508} {\bibfield  {journal} {\bibinfo
  {journal} {Phys. Rev. B}\ }\textbf {\bibinfo {volume} {84}},\ \bibinfo
  {pages} {140508} (\bibinfo {year} {2011})}\BibitemShut {NoStop}%
\bibitem [{\citenamefont {Sarma}(1963)}]{Sarma1963}%
  \BibitemOpen
  \bibfield  {author} {\bibinfo {author} {\bibfnamefont {G.}~\bibnamefont
  {Sarma}},\ }\bibfield  {title} {\bibinfo {title} {On the influence of a
  uniform exchange field acting on the spins of the conduction electrons in a
  superconductor},\ }\href
  {https://www.sciencedirect.com/science/article/pii/0022369763900076}
  {\bibfield  {journal} {\bibinfo  {journal} {Journal of Physics and Chemistry
  of Solids}\ }\textbf {\bibinfo {volume} {24}},\ \bibinfo {pages} {1029}
  (\bibinfo {year} {1963})}\BibitemShut {NoStop}%
\bibitem [{\citenamefont {Larkin}\ and\ \citenamefont
  {Ovchinnikov}(1965)}]{Larkin1964}%
  \BibitemOpen
  \bibfield  {author} {\bibinfo {author} {\bibfnamefont {A.~I.}\ \bibnamefont
  {Larkin}}\ and\ \bibinfo {author} {\bibfnamefont {Y.~N.}\ \bibnamefont
  {Ovchinnikov}},\ }\bibfield  {title} {\bibinfo {title} {{Nonuniform state of
  superconductors}},\ }\href@noop {} {\bibfield  {journal} {\bibinfo  {journal}
  {Sov. Phys. JETP}\ }\textbf {\bibinfo {volume} {20}},\ \bibinfo {pages} {762}
  (\bibinfo {year} {1965})}\BibitemShut {NoStop}%
\bibitem [{\citenamefont {Bobkova}\ and\ \citenamefont
  {Bobkov}(2014)}]{Bobkova2014}%
  \BibitemOpen
  \bibfield  {author} {\bibinfo {author} {\bibfnamefont {I.~V.}\ \bibnamefont
  {Bobkova}}\ and\ \bibinfo {author} {\bibfnamefont {A.~M.}\ \bibnamefont
  {Bobkov}},\ }\bibfield  {title} {\bibinfo {title} {Bistable state in
  superconductor/ferromagnet heterostructures},\ }\href
  {https://doi.org/10.1103/PhysRevB.89.224501} {\bibfield  {journal} {\bibinfo
  {journal} {Phys. Rev. B}\ }\textbf {\bibinfo {volume} {89}},\ \bibinfo
  {pages} {224501} (\bibinfo {year} {2014})}\BibitemShut {NoStop}%
\bibitem [{\citenamefont {Kittel}(2005)}]{Kittel2005}%
  \BibitemOpen
  \bibfield  {author} {\bibinfo {author} {\bibfnamefont {C.}~\bibnamefont
  {Kittel}},\ }\href@noop {} {\emph {\bibinfo {title} {{Introduction to solid
  state physics}}}},\ \bibinfo {edition} {8th}\ ed.\ (\bibinfo  {publisher}
  {Wiley},\ \bibinfo {year} {2005})\ p.\ \bibinfo {pages} {680}\BibitemShut
  {NoStop}%
\bibitem [{Dej(2015)}]{Dejene2015}%
  \BibitemOpen
  \bibfield  {title} {\bibinfo {title} {{Control of spin current by a magnetic
  YIG substrate in NiFe/Al nonlocal spin valves}},\ }\href
  {https://doi.org/10.1103/PHYSREVB.91.100404} {\bibfield  {journal} {\bibinfo
  {journal} {Physical Review B - Condensed Matter and Materials Physics}\
  }\textbf {\bibinfo {volume} {91}},\ \bibinfo {pages} {100404} (\bibinfo
  {year} {2015})}\BibitemShut {NoStop}%
\bibitem [{\citenamefont {Das}\ \emph {et~al.}(2019)\citenamefont {Das},
  \citenamefont {Dejene}, \citenamefont {{Van Wees}},\ and\ \citenamefont
  {Vera-Marun}}]{Das2019}%
  \BibitemOpen
  \bibfield  {author} {\bibinfo {author} {\bibfnamefont {K.~S.}\ \bibnamefont
  {Das}}, \bibinfo {author} {\bibfnamefont {F.~K.}\ \bibnamefont {Dejene}},
  \bibinfo {author} {\bibfnamefont {B.~J.}\ \bibnamefont {{Van Wees}}},\ and\
  \bibinfo {author} {\bibfnamefont {I.~J.}\ \bibnamefont {Vera-Marun}},\
  }\bibfield  {title} {\bibinfo {title} {{Temperature dependence of the
  effective spin-mixing conductance probed with lateral non-local spin
  valves}},\ }\href {https://doi.org/10.1063/1.5086423} {\bibfield  {journal}
  {\bibinfo  {journal} {Applied Physics Letters}\ }\textbf {\bibinfo {volume}
  {114}},\ \bibinfo {pages} {072405} (\bibinfo {year} {2019})},\ \Eprint
  {https://arxiv.org/abs/1812.09766} {arXiv:1812.09766} \BibitemShut {NoStop}%
\bibitem [{\citenamefont {Czeschka}\ \emph {et~al.}(2011)\citenamefont
  {Czeschka}, \citenamefont {Dreher}, \citenamefont {Brandt}, \citenamefont
  {Weiler}, \citenamefont {Althammer}, \citenamefont {Imort}, \citenamefont
  {Reiss}, \citenamefont {Thomas}, \citenamefont {Schoch}, \citenamefont
  {Limmer}, \citenamefont {Huebl}, \citenamefont {Gross},\ and\ \citenamefont
  {Goennenwein}}]{Czeschka2011}%
  \BibitemOpen
  \bibfield  {author} {\bibinfo {author} {\bibfnamefont {F.~D.}\ \bibnamefont
  {Czeschka}}, \bibinfo {author} {\bibfnamefont {L.}~\bibnamefont {Dreher}},
  \bibinfo {author} {\bibfnamefont {M.~S.}\ \bibnamefont {Brandt}}, \bibinfo
  {author} {\bibfnamefont {M.}~\bibnamefont {Weiler}}, \bibinfo {author}
  {\bibfnamefont {M.}~\bibnamefont {Althammer}}, \bibinfo {author}
  {\bibfnamefont {I.-M.}\ \bibnamefont {Imort}}, \bibinfo {author}
  {\bibfnamefont {G.}~\bibnamefont {Reiss}}, \bibinfo {author} {\bibfnamefont
  {A.}~\bibnamefont {Thomas}}, \bibinfo {author} {\bibfnamefont
  {W.}~\bibnamefont {Schoch}}, \bibinfo {author} {\bibfnamefont
  {W.}~\bibnamefont {Limmer}}, \bibinfo {author} {\bibfnamefont
  {H.}~\bibnamefont {Huebl}}, \bibinfo {author} {\bibfnamefont
  {R.}~\bibnamefont {Gross}},\ and\ \bibinfo {author} {\bibfnamefont
  {S.~T.~B.}\ \bibnamefont {Goennenwein}},\ }\bibfield  {title} {\bibinfo
  {title} {Scaling behavior of the spin pumping effect in ferromagnet-platinum
  bilayers},\ }\href {https://doi.org/10.1103/PhysRevLett.107.046601}
  {\bibfield  {journal} {\bibinfo  {journal} {Phys. Rev. Lett.}\ }\textbf
  {\bibinfo {volume} {107}},\ \bibinfo {pages} {046601} (\bibinfo {year}
  {2011})}\BibitemShut {NoStop}%
\bibitem [{\citenamefont {Meyer}\ \emph {et~al.}(2014)\citenamefont {Meyer},
  \citenamefont {Althammer}, \citenamefont {Geprägs}, \citenamefont {Opel},
  \citenamefont {Gross},\ and\ \citenamefont {Goennenwein}}]{Meyer2014}%
  \BibitemOpen
  \bibfield  {author} {\bibinfo {author} {\bibfnamefont {S.}~\bibnamefont
  {Meyer}}, \bibinfo {author} {\bibfnamefont {M.}~\bibnamefont {Althammer}},
  \bibinfo {author} {\bibfnamefont {S.}~\bibnamefont {Geprägs}}, \bibinfo
  {author} {\bibfnamefont {M.}~\bibnamefont {Opel}}, \bibinfo {author}
  {\bibfnamefont {R.}~\bibnamefont {Gross}},\ and\ \bibinfo {author}
  {\bibfnamefont {S.~T.~B.}\ \bibnamefont {Goennenwein}},\ }\bibfield  {title}
  {\bibinfo {title} {Temperature dependent spin transport properties of
  platinum inferred from spin hall magnetoresistance measurements},\ }\href
  {https://doi.org/10.1063/1.4885086} {\bibfield  {journal} {\bibinfo
  {journal} {Applied Physics Letters}\ }\textbf {\bibinfo {volume} {104}},\
  \bibinfo {pages} {242411} (\bibinfo {year} {2014})}\BibitemShut {NoStop}%
\end{thebibliography}%


\end{document}